\documentclass{article}

\usepackage[utf8]{inputenc} \usepackage[T1]{fontenc} \usepackage{geometry} \usepackage[english]{babel} \usepackage{graphicx} \usepackage{verbatim} \usepackage{amsmath}
\usepackage{amsfonts} \usepackage{bm}
\usepackage{siunitx}
\usepackage{xcolor}

\usepackage{natbib}
\usepackage{algorithm2e}

\usepackage{authblk}
\usepackage{hyperref}

\usepackage[linewidth=1pt,
middlelinecolor= black,
middlelinewidth=0.4pt,
roundcorner=1pt,
topline = false,
rightline = false,
bottomline = false,
rightmargin=0pt,
skipabove=0pt,
skipbelow=0pt,
leftmargin=-1cm,
innerleftmargin=1cm,
innerrightmargin=0pt,
innertopmargin=0pt,
innerbottommargin=0pt,
]{mdframed}

\usepackage{subfig}

\makeatletter
\newcommand{\removelatexerror}{\let\@latex@error\@gobble}
\makeatother

\newcommand{\RR}{\mathbb{R}}

\newcommand{\Id}{\text{Id}}

\newcommand{\Exp}{\operatorname{Exp}}

\newcommand{\Var}{\operatorname{Var}}

\graphicspath{{./fig/}}

\title{A model of brain morphological changes related to aging and Alzheimer's disease from cross-sectional assessments}

\author[1]{Raphaël~Sivera\footnote{Corresponding author at: Epione Research Project, INRIA Sophia-Antipolis, 2004, route des Lucioles, 06902 Sophia-Antipolis, France, raphael.sivera@inria.fr}}
\author[1]{Hervé~Delingette}
\author[1]{Marco~Lorenzi}
\author[1]{Xavier~Pennec}
\author[1]{Nicholas~Ayache}

\affil[ ]{for the  Alzheimer’s Disease Neuroimaging Initiative\footnote{Data used in preparation of this article were obtained from the Alzheimer’s Disease Neuroimaging Initiative (ADNI) database (adni.loni.usc.edu). As such, the investigators within the ADNI contributed to the design and implementation of ADNI and/or provided data but did not participate in analysis or writing of this report. A complete listing of ADNI investigators can be found at: \url{http://adni.loni.usc.edu/wp-content/uploads/how_to_apply/ADNI_Acknowledgement_List.p}}}
\affil[ ]{}
\affil[1]{Université Côte d'Azur, Inria Sophia Antipolis, Epione Research Project, France.}

\begin{document}

\maketitle

\section*{Abstract}

\label{sec:abstract}

In this study we propose a deformation-based framework to jointly model the influence of aging and Alzheimer's disease (AD) on the brain morphological evolution. Our approach combines a spatio-temporal description of both processes into a generative model. A reference morphology is deformed along specific trajectories to match subject specific morphologies. It is used to define two imaging progression markers: 1) a \emph{morphological age} and 2) a \emph{disease score}. 
These markers can be computed regionally in any brain region.

The approach is evaluated on brain structural magnetic resonance images (MRI) from the ADNI database.
The model is first estimated on a control population using longitudinal data, then, for each testing subject, the markers are computed cross-sectionally for each acquisition.
The longitudinal evolution of these markers is then studied in relation with the clinical diagnosis of the subjects and used to generate possible morphological evolutions.

In the model, the morphological changes associated with normal aging are mainly found around the ventricles, while the Alzheimer's disease specific changes are located in the temporal lobe and the hippocampal area. The statistical analysis of these markers highlights differences between clinical conditions even though the inter-subject variability is quite high. The model is also generative since it can be used to simulate plausible morphological trajectories associated with the disease.

Our method quantifies two interpretable scalar imaging biomarkers assessing respectively the effects of aging and disease on brain morphology, at the individual and population level. These markers confirm the presence of an accelerated apparent aging component in Alzheimer's patients but they also highlight specific morphological changes that can help discriminate clinical conditions even in prodromal stages. More generally, the joint modeling of normal and pathological evolutions shows promising results to describe age-related brain diseases over long time scales.

Keywords: aging, Alzheimer's disease, brain morphology, deformations, spatio-temporal model, imaging biomarkers.

\section{Introduction}

Age-related diseases are a growing public health concern with the aging of the population. For this reason, a precise description of aging would be useful to predict and describe the evolution of these diseases. In complement to the \emph{chronological age}, i.e. the time elapsed since birth, one would like to estimate a \emph{biological age} that reflects the current physiological, functional or structural status of an organ relatively to the aging changes. However there is no unified way to describe aging in a clinical context since aging is a complex process which affects every part of the body with specific mechanisms and specific rate. As a consequence multiple theories of aging have been proposed~\citep[][]{medvedev_attempt_1990}, leading to the definition of surrogate age variables based on the quantification of biological changes.

\subsection{Modeling brain morphological aging}
In this paper we focus on the aging of the brain based on the study of its shape evolution. The brain is not exempt from aging and a decline of cognitive processing speed, working memory, inhibitory function, and long-term memory is generally observed.
This decline has been associated with neural activity changes~\citep{park_adaptive_2009} and it was also shown to be directly correlated with structural changes such as brain atrophy, cortex thinning and decrease of white matter integrity~\citep{rosen2003differential,rodrigue2004shrinkage}.

\begin{figure}[ht!]
  \centering
    \def\svgwidth{0.6\textwidth}
  \begingroup  \makeatletter  \providecommand\color[2][]{    \errmessage{(Inkscape) Color is used for the text in Inkscape, but the package 'color.sty' is not loaded}    \renewcommand\color[2][]{}  }  \providecommand\transparent[1]{    \errmessage{(Inkscape) Transparency is used (non-zero) for the text in Inkscape, but the package 'transparent.sty' is not loaded}    \renewcommand\transparent[1]{}  }  \providecommand\rotatebox[2]{#2}  \ifx\svgwidth\undefined    \setlength{\unitlength}{662.45331051bp}    \ifx\svgscale\undefined      \relax    \else      \setlength{\unitlength}{\unitlength * \real{\svgscale}}    \fi  \else    \setlength{\unitlength}{\svgwidth}  \fi  \global\let\svgwidth\undefined  \global\let\svgscale\undefined  \makeatother  \begin{picture}(1,0.65947759)    \put(0,0){\includegraphics[width=\unitlength,page=1]{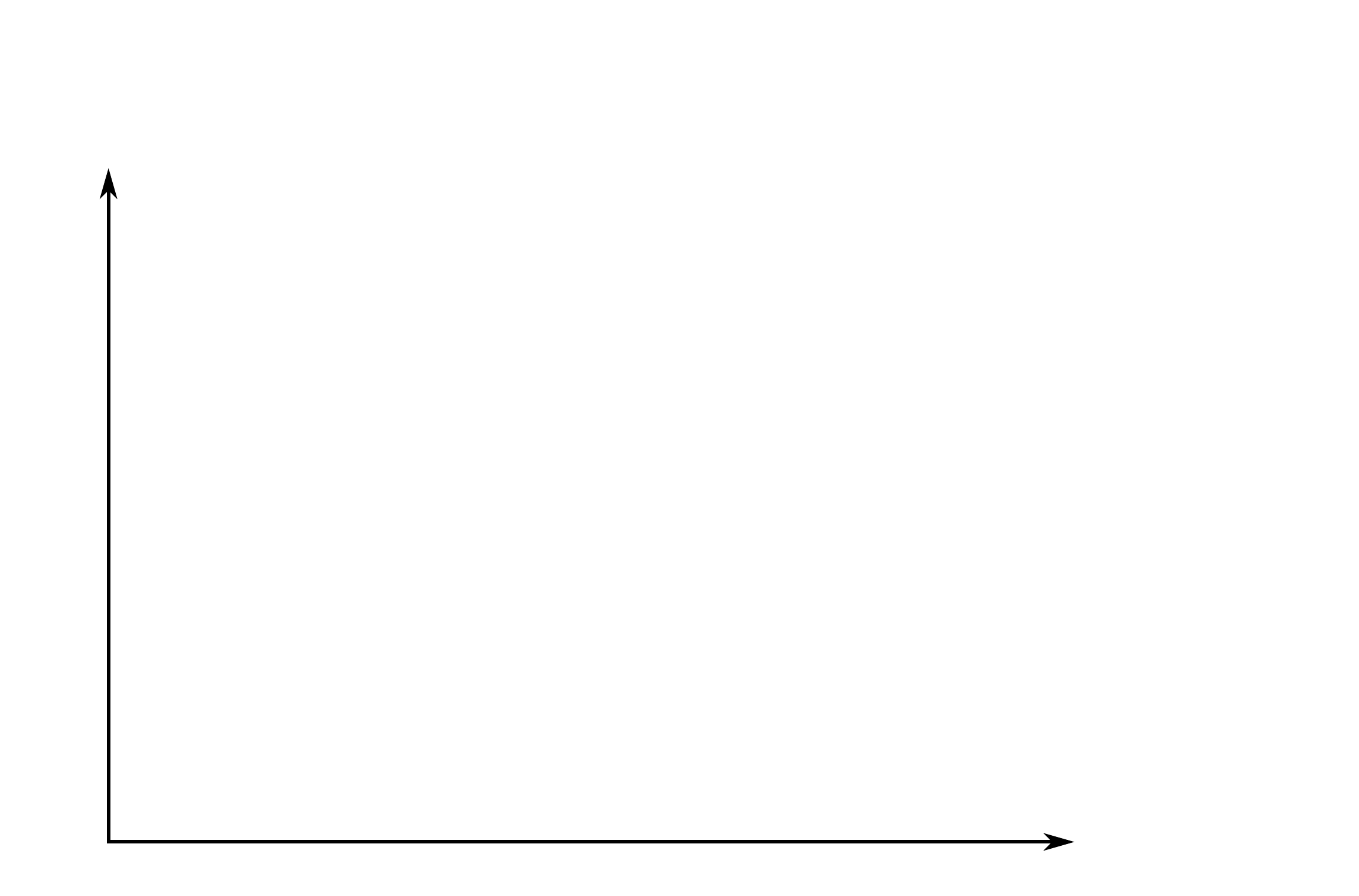}}    \put(0.58485506,0.00722179){\color[rgb]{0,0,0}\makebox(0,0)[lb]{\smash{chronological age}}}    \put(-0.00208517,0.53816439){\color[rgb]{0,0,0}\makebox(0,0)[lb]{\smash{morphological age}}}    \put(0,0){\includegraphics[width=\unitlength,page=2]{schema_idea_origin.pdf}}    \put(0.70,0.60){\color[rgb]{0.6275,0.2745,0}\makebox(0,0)[lt]{\begin{minipage}{0.3137101\unitlength}\raggedright Evolution of a subject with AD\end{minipage}}}    \put(0.74,0.44){\color[rgb]{0,0.31372549,0}\makebox(0,0)[lt]{\begin{minipage}{0.29289742\unitlength}\raggedright Evolution of a healthy subject\end{minipage}}}  \end{picture}\endgroup   \caption{Schematic representation of two evolutions relative to an hypothetical \emph{morphological age} reflecting the structural status of the brain relative to the aging process.}
  \label{fig:schema_idea}
\end{figure}

The normal brain morphology has been studied in image-based studies from the development stage to the most advanced ages. Measurements of brain structures (volumes, cortical thickness, etc.) have been performed for wide age ranges and the statistical analysis of the evolution of these measurements helps in providing an initial understanding of the normal brain shape evolution across life span~\citep{good_voxel-based_2001,long_healthy_2012}. These descriptions have been used to estimate models characterizing brain aging in order to highlight differences across brain areas~\citep{hutton_comparison_2009,sowell_mapping_2003}. 
The inverse problem, i.e. how to associate an age to a brain image, was also addressed. Models have been designed to estimate the \emph{chronological age}~\citep{cole_brain_2017} from an image but they can also be used to characterize abnormal evolutions. For instance, a mean brain age gap estimate was highlighted for Alzheimer's patients~\citep{franke_estimating_2010}. More generally, these surrogate brain age estimates have been associated with an increase of risk factors for several age-related disorders such as cardio-vascular diseases~\citep{carli_measures_2005, velsen_brain_2013}. In a longitudinal setting, a brain age measurement could be used to compare the evolution of several clinical conditions (see Figure~\ref{fig:schema_idea}).

\subsection{Disease progression modeling}
Brain aging is often associated with the development of neurodegenerative pathologies. For example, it is estimated that one in three people over 85 have Alzheimer's disease (AD), the most common form of dementia~\citep{association_2017_2017}. This disease comes with its own specific apparent brain morphological changes~\citep{ohnishi2001changes} and computer-aided diagnosis techniques using neuroimaging features have shown promising results to classify and to predict clinical evolutions~\citep{davatzikos_longitudinal_2009, kloppel_diagnostic_2012, schmitter_evaluation_2015}. 
Longitudinal studies provide us with multiple acquisitions at different times for every subject but the disease affects patients over several decades, starting even before the first symptoms occur, and few studies follow a significant number of subjects over such long times. Progression models have been developed to describe the global evolution and to put in relation the individual trajectories that could only be observed a limited number of times. They have been used to model the progression of biomarkers~\citep{fonteijn_event-based_2012, donohue_estimating_2014} but also directly brain shape~\citep{cury_spatio-temporal_2016} or spatio-temporal patterns in brain images~\citep{koval_statistical_2017, schiratti_bayesian_2016}. These models produce good representations of the disease progression and can combine a variety of available biomarkers for patient monitoring~\citep{bilgel_multivariate_2016, lorenzi2017probabilistic}.

\subsection{Toward a joint model of brain aging and disease progression}
\label{sec:subsec:intro_toward_joint_model}
The morphological aging of the brain and disease progression have generally been modeled independently. However we know that the structural features used in diagnosis (e.g cortical thickness or atrophy patterns) are also generally related to age. Indeed aging and neurodegenerative diseases involve intertwined processes with entangled consequences. Surrogate age measurements have been used to support the disease characterization~\citep{franke_longitudinal_2012} or to put aside the aging part in order to focus on the disease specific changes.
\citet{lorenzi_disentangling_2015} proposed to model the normal aging evolution to separate the contribution of aging from a remainder that is not explicitly modeled. This remainder is then used to describe the pathological evolution. However, this method does not propose an intrinsic model of the disease progression making it difficult to describe and to characterize disease specific changes.

In this study, we propose a generative model of the brain morphological evolution that jointly takes into account the normal aging and the disease effects. Our model is based on the approach proposed by~\citet{lorenzi_disentangling_2015} and gives a deformation-based description of subject trajectories.
It extends the original approach by explicitly modeling the disease specific brain morphological evolution. In addition to the apparent \emph{morphological age} computed in the proposed approach, it allows us to get a \emph{disease score}, thus providing two morphological imaging biomarkers accounting for the progression of the two main ongoing processes: normal aging and Alzheimer's disease. These biomarkers are estimated cross-sectionnally from a single structural MRI but the model estimation exploits longitudinal data to match as accurately as possible the morphological evolution of the subjects.

We introduce in section~\ref{sec:method:model} the generative model used to represent the brain morphology. Section~\ref{sec:method:estimation} focuses on the estimation of the model parameters and on the inverse problem solved to compute the morphological age and the disease score of a subject. Experimental results are then presented in section~\ref{sec:results} in order to evaluate our model and parameter estimation procedure. We illustrate how the model helps to describe the evolution of subjects at different disease stages using the ADNI database. Then we show how the two proposed markers can help to follow the evolution of elderly patients.
Finally, limitations and perspectives are discussed in section~\ref{sec:discussion}.

\section{Definition of the generative model}
\label{sec:method:model}

In the sequel, we quantify differences between morphologies by spatial deformations that can be estimated from magnetic resonance images (MRI) through non-linear image registration. A deformation represents either a morphological difference between the anatomies of two subjects or a longitudinal evolution of one subject-specific anatomy. Therefore deformation based frameworks are well suited to define a parametric model of the morphology. Our approach involves a population template morphology which is parametrized by two progression markers: the morphological age and the disease score.

In section~\ref{subsec:method:reference_space}, we expose the main ideas behind the definition of our reference parametric space. Then in section~\ref{subsec:method:morphological_evolution}, we explain how the morphological evolutions are modeled in our framework. Finally in  section~\ref{subsec:method:individual_morphology}, we show how we use the deformation framework to model individual morphologies relatively to this reference.

\subsection{A space of reference morphologies}
\label{subsec:method:reference_space}

In deformation based morphometry, a single morphology is classically used to approximate and represent a population. For a set of images $\{I_k\}$,  we define a common reference image $T_0$, called template. The difference of morphology between the subject $k$ and the template is modeled with a spatial deformation $\phi_k$ and an intensity noise $\epsilon_k$ is added in the subject space accounting for local intensity variability. Therefore the images are modeled as follows:
$$I_k = T_0 \circ \phi_k + \epsilon_k.$$

In our approach, we want to take into account two major processes that affect the brain morphology over time: the normal aging and the disease evolution. To do so, we model the effects of these processes on the template using a deformation $\Phi$ parametrized by two variables that measure the progress of each process: the \emph{morphological age} $\lambda_{MA}$ and the \emph{disease score} $\lambda_{DS}$.
The two variables $\lambda_{MA}$ and $\lambda_{DS}$ can be seen as time variables and are scaled to correspond to years of standard evolution. In this model, $T_0 \circ \Phi(\lambda_{MA}, \lambda_{DS})$ represents the template morphology after $\lambda_{MA}$ years of normal aging and $\lambda_{DS}$ years of normalized disease progression.

In the ideal case, the morphological age is equal to the chronological age. Therefore, $T_0 \circ \Phi(t, 0)$ represents the morphology of a $t$ years old healthy subject. Similarly $T_0 \circ \Phi(t, t')$ would be the typical morphology of a AD patient of age $t$ with a disease duration of $t'$ years. If one is able to associate an age $\lambda_0$ to the image $T_0$, and assumes that this is the image of a healthy subject, then it is natural to enforce $\Phi(\lambda_0, 0) = \Id$. 

The parametric subspace of images generated like this will be used as a reference:
 \begin{align}
   \mathcal{T} = \left\{ T_0 \circ \Phi(\lambda_{MA}, \lambda_{DS}) \text{ for } \lambda_{MA}, \lambda_{DS} \in \RR \right\}.
 \end{align}

An initial model of the subject images is then:
$$I_k = T_0 \circ \Phi(\lambda_{MA}^k, \lambda_{DS}^k) \circ \phi_k + \epsilon_k,$$
where $\lambda_{MA}^k$ and $\lambda_{DS}^k$ are the subject morphological age and disease score, while $\phi_k$ encodes differences specific to the subject morphology. Of course multiple options are available to combine the longitudinal deformation $\Phi$ with the subject specific changes $\phi^k$. Here we write this operation as a right-composition but this choice will be discussed in section~\ref{subsec:method:morphological_evolution} and a similar but symmetrical operation will later be used.

Also, in the reference space, trajectories parametrized by time $ t \mapsto T_0 \circ \Phi({\lambda_{MA}(t)}, {\lambda_{DS}(t)}) $ give possible morphological evolutions where the morphological age and the disease score can be seen as reparametrization of time.
In particular, it defines two archetypal trajectories (\emph{i.e.} ideal models of evolution): the \emph{normal aging template trajectory} $t \mapsto T_0 \circ \Phi(t, 0)$ and the \emph{disease specific template trajectory} $t \mapsto T_0 \circ \Phi(\lambda_0, t)$.

Here, we assume that each progression of the two major processes (aging and disease) can be described with only one parameter. This implies that the evolution of healthy aging is similar for each subject, following the normal aging template trajectory, even if the speed of aging may vary from one subject to the other. Similarly, the disease progression is described using a single trajectory and we combine both template trajectories to model pathological evolutions.

\subsection{Modeling the morphological evolution of the template}
\label{subsec:method:morphological_evolution}

In a simplified approach, the template trajectories are assumed to be geodesics in an appropriate deformation space. Geodesics define continuous paths that can be easily parametrized and constrained to allow regularity in time. They can be used to interpolate between two anatomies or to approximate more complex trajectories~\citep{christensen_3d_1994, wang_large_2007}. 
In this paper, we use the stationary velocity field (SVF) framework~\citep{arsigny_log-euclidean_2006} for its ability to describe complex and realistic diffeomorphic (smooth and invertible) brain deformations in a straightforward manner~\citep{lorenzi_lcc-demons:_2013}. In this framework, the observed anatomical changes are encoded by diffeomorphisms which are parametrized with the flow of SVFs. Within this setting, the metric between deformations is not chosen \emph{a priori} even if we need a regularization criterion for the registration. To compute a deformation $\phi$ we integrate trajectories along the vector field $v$ for a unit of time. 
$$ \phi(x) = \phi_1(x) = \int^1_0{v(\phi_t(x)) dt} \text{\qquad  with \qquad} \phi_0 = \Id. $$ 
This relationship is denoted as the group exponential map $\phi = \Exp(v)$.

By writing $\Phi(\lambda_{MA}, \lambda_{DS}) = \Exp(\lambda_{MA} \mathrm{v}_{A} + \lambda_{DS}\mathrm{v}_{D})$, we propose a linear model in the SVF space (\emph{i.e.} the space of the parameter of the deformations)  parametrized by two SVFs $\mathrm{v}_{A}$ and $\mathrm{v}_{D}$. In particular, the two template trajectories are then separately parametrized: $\mathrm{v}_{A}$ controls the normal aging template trajectory and $\mathrm{v}_{D}$ the disease specific template trajectory.

For each subject, the processes are meant to be intertwined and this can be modeled in different ways depending on the parametrization of the trajectories, for instance a right or a left composition. The proposed linear combination of the parameters provides us a middle ground. Indeed in the SVF setting, the relationship between composition and the linear combination of SVFs is given by the Baker-Campbell-Hausdorff formula~\citep{bossa_contributions_2007} and the linear combination of the SVFs is equivalent to alternate between right and left composition with infinitesimal steps.

 To sum up, the longitudinal deformation $\Phi$ modeling the effects of aging and the disease on a reference morphology $T_0$ is parametrized by two SVFs: $\mathrm{v}_{A}$ and $\mathrm{v}_{D}$. This ideal model generates a surface $\mathcal{T}$ of possible images describing the evolution of the template morphology:
 
 \begin{align}
   \mathcal{T} = \left\{ T_0 \circ \Exp(\lambda_{MA} \mathrm{v}_{A} + \lambda_{DS}\mathrm{v}_{D}) \text{ for } \lambda_{MA}, \lambda_{DS} \in \RR \right\}. 
 \end{align}

\subsection{Individual morphological variability and generative model}
\label{subsec:method:individual_morphology}

An individual image is modeled as follows:
 \begin{align}
   I_k = T_0 \circ \Exp((\lambda_{MA}^k - \lambda_0) \mathrm{v}_{A} + \lambda_{DS}^k \mathrm{v}_{D}) \circ \phi_k + \epsilon_k,
 \end{align}
where the choice of the intensity noise $\epsilon_k$ is implicitly related to the registration similarity metric. To specify the constraint on $\phi_k$, we define a subject specific residual SVF $w^k_{r}$ ($r$ stands for residual) such that:
$$\Exp((\lambda_{MA}^k - \lambda_0) \mathrm{v}_{A} + \lambda_{DS}^k \mathrm{v}_{D}) \circ \phi_k = \Exp((\lambda_{MA}^k - \lambda_0) \mathrm{v}_{A} + \lambda_{DS}^k \mathrm{v}_{D} + w^k_{r}).$$
In this formula, $\Exp(w^k_{r})$ is approximately equal to $\phi_k$ given the first order of the BCH equation between composition and linear combination of SVFs.
Moreover, we wish to have the subject specific deformation to encode what cannot be described using the template trajectories. That is why we impose $w^k_{r}$ to be orthogonal to both $\mathrm{v}_{A}$ and $\mathrm{v}_{D}$.

\begin{figure}[ht!]
  \centering
    \def\svgwidth{0.7\textwidth}
  \begingroup  \makeatletter  \providecommand\color[2][]{    \errmessage{(Inkscape) Color is used for the text in Inkscape, but the package 'color.sty' is not loaded}    \renewcommand\color[2][]{}  }  \providecommand\transparent[1]{    \errmessage{(Inkscape) Transparency is used (non-zero) for the text in Inkscape, but the package 'transparent.sty' is not loaded}    \renewcommand\transparent[1]{}  }  \providecommand\rotatebox[2]{#2}  \ifx\svgwidth\undefined    \setlength{\unitlength}{538.69189165bp}    \ifx\svgscale\undefined      \relax    \else      \setlength{\unitlength}{\unitlength * \real{\svgscale}}    \fi  \else    \setlength{\unitlength}{\svgwidth}  \fi  \global\let\svgwidth\undefined  \global\let\svgscale\undefined  \makeatother  \begin{picture}(1,0.63904587)    \put(0,0){\includegraphics[width=\unitlength,page=1]{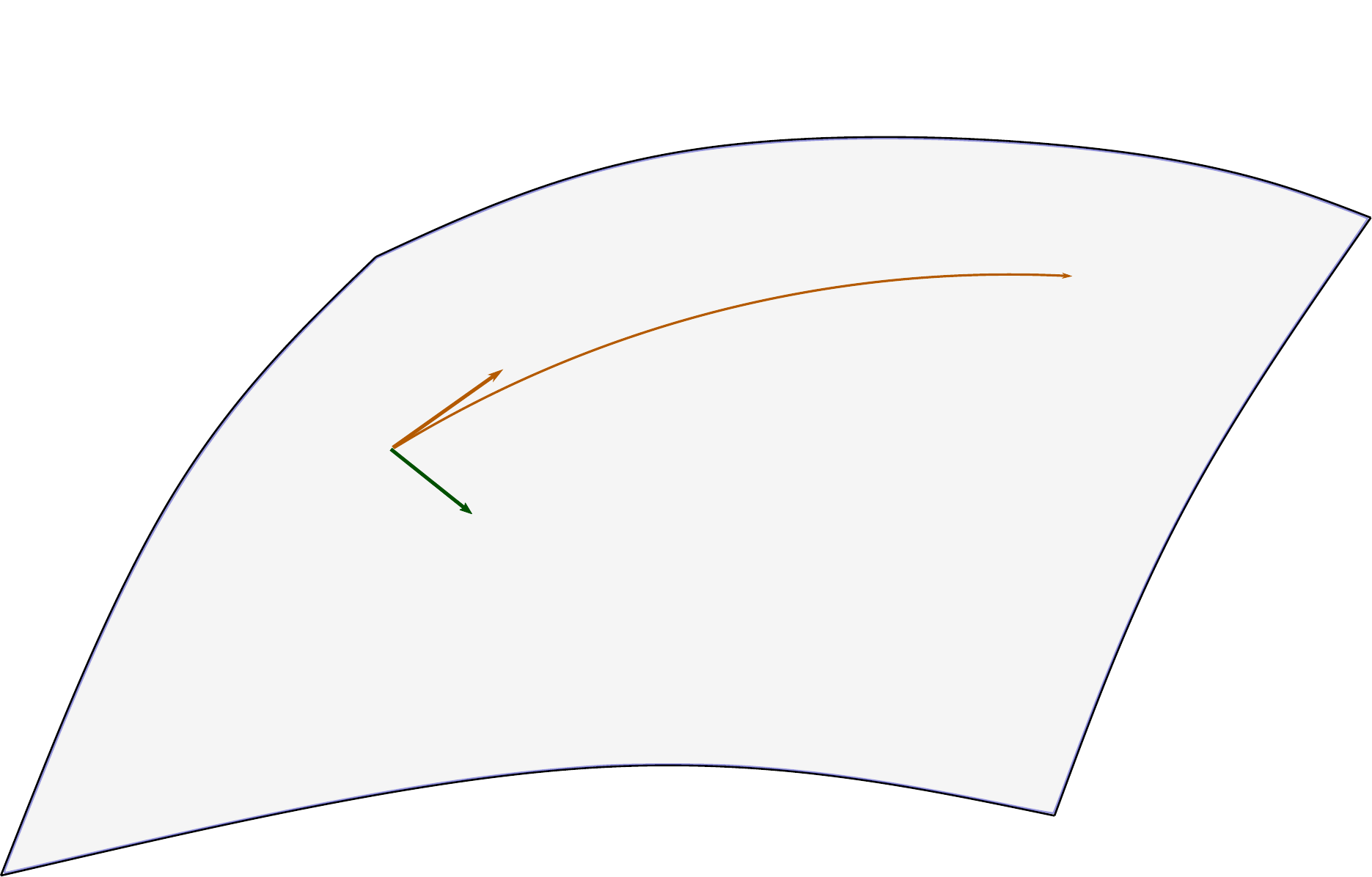}}    \put(0.3150998,0.36794829){\color[rgb]{0.70588235,0.35294118,0}\rotatebox{-2.4585305}{\makebox(0,0)[lb]{\smash{$\mathrm{v}_{D}$}}}}    \put(0.31,0.25){\color[rgb]{0,0.31372549,0}\rotatebox{0.89999987}{\makebox(0,0)[lb]{\smash{$\mathrm{v}_{A}$}}}}    \put(0,0){\includegraphics[width=\unitlength,page=2]{schema_parameters_cross_only_courbe.pdf}}    \put(0.65462361,0.30662072){\color[rgb]{0,0,0}\makebox(0,0)[lb]{\smash{ }}}    \put(0,0){\includegraphics[width=\unitlength,page=3]{schema_parameters_cross_only_courbe.pdf}}
    \put(0,0){\includegraphics[width=\unitlength,page=4]{schema_parameters_cross_only_courbe.pdf}}    \put(0.40543726,0.40556639){\color[rgb]{0.70588235,0.35294118,0}\makebox(0,0)[lb]{\smash{$\lambda_{DS}^k$}}}    \put(0.42467667,0.20053636){\color[rgb]{0,0.31372549,0}\makebox(0,0)[lb]{\smash{$\lambda_{MA}^k$}}}    \put(0.47693977,0.44659465){\color[rgb]{0,0,0.64705882}\makebox(0,0)[lb]{\smash{$w_r^k$}}}    \put(0.05570275,0.13){\color[rgb]{0.35294118,0.35294118,0.47058824}\makebox(0,0)[lt]{\begin{minipage}{0.30291619\unitlength}\raggedright \textit{non linear subspace of  images }  $\mathcal{T}$\end{minipage}}}    \put(0.63399778,0.45065351){\color[rgb]{0.70588235,0.35294118,0}\makebox(0,0)[lb]{\smash{t $\mapsto T_0 \circ \Exp(t \mathrm{v}_{D})$}}}    \put(0.57852632,0.22350408){\color[rgb]{0.13333333,0.33333333,0}\makebox(0,0)[lb]{\smash{t $\mapsto T_0 \circ \Exp(t \mathrm{v}_{A})$}}}    \put(0.63399778,0.40065351){\color[rgb]{0.70588235,0.35294118,0}\makebox(0,0)[lb]{\smash{AD-specific evolution}}}    \put(0.57852632,0.17350408){\color[rgb]{0.13333333,0.33333333,0}\makebox(0,0)[lb]{\smash{Healthy aging}}}

    \put(0.425,0.59){\color[rgb]{0,0,0}\makebox(0,0)[lb]{\smash{observed anatomy}}}    \put(0.432,0.56){\color[rgb]{0,0,0}\makebox(0,0)[lb]{\smash{$I_k$}}}  \end{picture}\endgroup   \caption{Our two-trajectory model. The template image $T_0$, the normal aging template trajectory parametrized by $\mathrm{v}_{A} $ and the disease specific template trajectory parametrized by $\mathrm{v}_{D}$ define a subspace of possible morphologies of reference. An individual morphology is characterized by a morphological age $\lambda_{MA}^k$, a disease score  $\lambda_{DS}^k$, and an SVF $w^k_{r}$ modeling the subject-specific part. Each image can be projected onto the template subspace $\mathcal{T}$ using a decomposition of the deformation between the image and the template $T_0$.}
  \label{fig:schema_model}
\end{figure}

As we can see in Figure~\ref{fig:schema_model}, the model parametrized by $\mathrm{v}_{A}$ and $\mathrm{v}_{D}$ allows us to characterize the subject morphology with two scalar variables, the morphological age $\lambda_{MA}^k$ and the disease score $\lambda_{DS}^k$, and a SVF $w^k_{r}$ for the subject-specific part. The orthogonality constraint makes the description of the subject uniquely defined. We denote by $w^k$ the subject-to-template deformation SVF:

\begin{align}
  w^k = (\lambda_{MA}^k - \lambda_0) \mathrm{v}_{A} + \lambda_{DS}^k \mathrm{v}_{D} + w^k_{r}.
\end{align}

\section{Estimation of the model parameters}
\label{sec:method:estimation}

The model parameters are of two kinds: the population parameters ($T_0$, $\mathrm{v}_{A}$ and $\mathrm{v}_{D}$) and the subjects parameters ($\lambda_{MA}^{k}$, $\lambda_{DS}^{k}$ and $w^{k}_{r}$).
To tackle this joint estimation problem in a computationally efficient way, several assumptions are made:
\begin{enumerate}
\item When available, the longitudinal evolution in the template space can be approximated by the transported deformation estimated in the subject space. Parallel transport algorithms are commonly used in the geometrical analysis of longitudinal data. The use of geodesic parallelograms is in general an efficient way to bring individual trajectories in a common reference space~\citep{lorenzi_geodesics_2013}. In practice, it allows us to work only with intra-subject deformation to estimate the model population parameters. It simplifies the optimization and is also more stable as the intra-subject variability is in general smaller than the inter-subject one.
\item We also assume, while estimating the population parameters, that the aging speed and the disease progression speed are constant for all the subjects in the training set. 
\item Intra-subject deformations are relatively small and smooth. Consequently, the registration regularization has less impact on the estimated deformation. This consideration allows us to estimate these longitudinal evolutions independently of the population model. 
\end{enumerate}

These assumptions allow us to efficiently decompose the problem of the parameter estimation. First, subjects with longitudinal data are processed independently and the intra-subject evolutions are modeled in the subject space. Then the population parameters ($T_0$, $\mathrm{v}_{A}$ and $\mathrm{v}_{D}$) are estimated using only intra-subject longitudinal evolutions. Finally, the subjects' parameters are estimated.

\subsection{Estimation of the template trajectories $\mathrm{v}_{A}$ and $\mathrm{v}_{D}$ in a given template space}
\label{sec:subsec:vavd}

In this section we suppose that we know $T_0$ and that we can compute the subject-to-template deformation $w^k$ for a reference time point. We also consider that we have longitudinal data for every subject. 

First we address the inverse problem of estimating the intra-subject evolution parameters with the framework proposed by~\citet{hadj-hamou_longitudinal_2016}.
Images are preprocessed, rigidly aligned to the MNI-152 template and then longitudinally registered. Intra-subject deformations between follow-up images and the baseline image are computed using non-linear registration. The resulting intra-subject model in the subject's space is estimated using ordinary least square regression in the tangent space of SVFs. It is equivalent to the assumption that the deformation noises are centered, uncorrelated and have equal variance in the space of SVFs.

Then for a given template $T_0$ and subject-to-template deformation $w^k$, the intra-subject model can be transported using parallel transport in the template space to get $v^k$. This deformation can be decomposed along the template trajectories giving a morphological aging rate (noted $s_{MA}^k$), a disease progression rate (noted $s_{DS}^k$) and an orthogonal component (noted $v_{r}^k$):

$$v^k = s_{MA}^k \mathrm{v}_{A} + s_{DS}^k \mathrm{v}_{D} + v_{r}^k. $$

These progression rates are strongly related to the evolution of the morphological age and of the disease score (in a ideal euclidean case $s_{MA}^k = d \lambda_{MA}^{k} / dt$ and  $s_{DS}^k = d \lambda_{DS}^{k} / dt$) and they will be used to normalize the speed of evolution.
The estimation is done on two groups of subjects: a group $\mathcal{G}_h$ composed by healthy subjects and a group $\mathcal{G}_{ad}$ composed by patients diagnosed with AD. We assume that each healthy subject of $\mathcal{G}_h$ is aging at normal speed $s_{MA}^{k} = 1, \forall k \in \mathcal{G}_h$ and does not have any evolution toward the disease $s_{DS}^{k} = 0, \forall k \in \mathcal{G}_h$. Similarly, each patient of $\mathcal{G}_{ad}$ has a normal morphological aging rate $s_{MA}^k = 1, \forall k \in \mathcal{G}_{ad}$ and a constant unit disease progression rate $s_{DS}^{k} = 1, \forall k \in \mathcal{G}_{ad}$.
Finally the subject specific components are assumed to be centered, uncorrelated and to have a fixed variance. The maximum likelihood problem writes:
\begin{align}
\min_{\mathrm{v}_{A}, \mathrm{v}_{D}}{\sum_{k \in \mathcal{G}_h}{\| v^k - \mathrm{v}_{A} \|^2} + \sum_{k \in \mathcal{G}_{ad}}{\| v^k - \mathrm{v}_{A} - \mathrm{v}_{D} \|^2} }.
\end{align}

The solution of the optimization problem is explicit:
\begin{align}
  \hat{\mathrm{v}}_{A} &= \frac{1}{|\mathcal{G}_h|} \sum_{k \in \mathcal{G}_h}{v^k}, \\
  \hat{\mathrm{v}}_{D} &=  \left(\frac{1}{|\mathcal{G}_{ad}|} \sum_{k \in \mathcal{G}_{ad}}{v^k}\right) - \hat{\mathrm{v}}_{A}.
\end{align}
 We should however note that $\| \hat{\mathrm{v}}_{A} \| $ (resp. $\| \hat{\mathrm{v}}_{D} \| $) is a biased estimator of $\| \mathrm{v}_{A} \| $ (resp. $\| \mathrm{v}_{D} \| $). We detail the bias estimation in the Appendix~\ref{subsec:appendix:bias}.

\subsection{Estimation of the template morphology $T_0$}

The population specific template morphology is computed using the algorithm proposed by~\citet{guimond_average_2000} by alternating the registration of subject images to the template and the recomputation of the template intensities. However, in our approach, the subjects' images do not need to be registered to $T_0$ directly but to their projection on the template space. 
To tackle this problem, we propose an iterative procedure where we register the image to its current projection on the reference space. Algorithm~\ref{algo:iterative_registration} details the procedure with simplified notations in the general case where $w$ parametrizes the deformation between the image $I$ and $T$. The reference linear subspace of SVF is denoted $\mathcal{T}$ and $w$ is decomposed accordingly $w = w_t + w_r$ with $w_t \in \mathcal{T}, w_r \in \mathcal{T^{\perp}}$. The registration regularization should only be applied to the residual part $w_{r}$. In the context of the LCC-demons registration algorithm, it boils down to the following minimization problem~\citep[see][]{lorenzi_lcc-demons:_2013}:
$$ \min_{w_t \in \mathcal{T}, w_r \in \mathcal{T^{\perp}}, w'} \text{Sim}(I, T, \Exp(w_t + w_r)) + \text{Dist}(w_r, w') + \text{Reg}(w'). $$
The idea is to alternate between the optimization and the projection on the constraints.

\begin{figure}[!t]
  \removelatexerror
  \begin{algorithm}[H]
    \KwData{an image $I$, a template image $T$ and linear space of SVF $\mathcal{T}$}
    \KwResult{two SVFs: $w_t \in \mathcal{T}, w_r \in \mathcal{T^{\perp}}$}
    $w_t=0$ \;
    \Repeat{ $ w_r \perp \mathcal{T} $ }{
      $w_r = \text{registration}(T \circ \Exp(w_t), I)$ \;
      $w_t = w_t + \text{proj}_{\mathcal{T}}(w_r) $\;
    }
    \caption{Iterative registration algorithm}
    \label{algo:iterative_registration}
  \end{algorithm}
\end{figure}

Since we do not have any theoretical guarantee on the convergence of the algorithm, the stability and the convergence will be evaluated empirically. 
As the template estimation also involves iterative search, we can combine both algorithms for a faster optimization.
The projection coordinates are kept from one iteration to the next and the images are registered to their estimated projections in the template space. The deformation update $u$ is then computed and finally the new atlas image $T$ and the estimated projections are updated (see Algorithm~\ref{algo:iterative_template}).

\begin{figure}[!t]
  \removelatexerror
  \begin{algorithm}[H]
    \KwData{a set of images $(I_k)$ and a linear space of SVF $\mathcal{T}$}
    \KwResult{a template image $T$, a set of pairs of SVFs $(w_t^k, w_r^k)$}
    $w_t^k=0$ for all $k$\;
    initialize $T$ \; 
    \Repeat{ convergence }{
      $w_r^k = \text{registration}(T \circ \Exp(w_t^k), I_k)$ for all $k$\;
      $u = \text{mean}( w_r^k ) $ \;
      $T = \text{mean}( I_k \circ \Exp(- w_r^k + u))$ \;
      $w_t^k = w_t^k + \text{proj}_{\mathcal{T}}(w_r^k - u) $ for all $k$\;
    }
    \caption{Iterative template space estimation algorithm}
    \label{algo:iterative_template}
  \end{algorithm}
\end{figure}

In this work, the intra-subject models are transported to the template space to update the template trajectories parametrizing $\mathcal{T}$ at each iteration (using the approach described in the previous section~\ref{sec:subsec:vavd}). $T$ is initialized using the MNI-152 template and the convergence is manually assessed comparing the template for successive iterations. At convergence, we obtain the template image $T_0$ and both template trajectories $\mathrm{v}_{A}$ and $\mathrm{v}_{D}$.

\subsection{Estimation of the subject's parameters}

When the population parameters are learned, the estimation of the individual parameters for a new subject is relatively simple. The deformation $w^k$ is computed by registration between a subject image and the template using Algorithm~\ref{algo:iterative_registration}, and then linearly decomposed, $w^k = (\lambda_{MA}^k - \lambda_0) \mathrm{v}_{A} + \lambda_{DS}^k \mathrm{v}_{D} + w^k_{r}$, by solving the following linear system:
\begin{align}
  w^k \cdot \mathrm{v}_{A} &= \|\mathrm{v}_{A}\|^2 (\lambda_{MA}^k - \lambda_0)               +  \mathrm{v}_{D} \cdot \mathrm{v}_{A} \lambda_{DS}^k, \\
  w^k \cdot \mathrm{v}_{D} &= \mathrm{v}_{D} \cdot \mathrm{v}_{A} (\lambda_{MA}^k - \lambda_0) +  \|\mathrm{v}_{D}\|^2 \lambda_{DS}^k.
  \label{eqt:subject_parameters}
\end{align}

In practice, the estimation is not exact because we work with the noisy estimators $\hat{\mathrm{v}}_{A}$ and $\hat{\mathrm{v}}_{D}$.
The linear decomposition can also be computed locally or by using any voxel weighting scheme for the scalar product. When longitudinal data is available, this estimation is independently done for each time point.

\section{Results}
\label{sec:results}

\subsection{Experiments with synthetic data}
\label{subsec:res:synthetic_data}

We first evaluate our approach using synthetic data in order to assess the accuracy and the reproducibility of the biomarkers estimation. Realistic longitudinal MRIs are simulated using the software proposed by \citet{khanal_biophysical_2016}. The simulation algorithm relies on a biophysical model of brain deformation and can be used to generate longitudinal evolutions with specific atrophy patterns. In this context, local atrophy is measured by the divergence of the stationary velocity field.

\subsubsection{Simulated dataset}

In this controlled experiments we choose to simulate two populations that are characterized by their atrophy patterns and that respectively emulate healthy controls and AD patients. Atrophy of the aging brain and the effect of AD have been extensively studied~\citep{pini_brain_2016} and the atrophy measurements may vary depending on the methodology and the population studied.
In this experiment, we choose to prescribe piecewise-constant atrophy map with constant value in brain areas delimited by the segmentation provided by \emph{FreeSurfer}~\citep{fischl2002whole}. For every subject, the atrophy value of a region is sampled around a fixed population mean with an additive Gaussian noise of relative standard deviation of 5\% .
The healthy population is designed to have a small atrophy in the whole brain while the patients have a stronger atrophy especially in the hippocampal areas and the temporal poles. The means are chosen to give the order of magnitude of a one year evolution accordingly to what was reported in~\citet{fjell_brain_2010} for healthy aging and in~\citet{carmichael_coevolution_2013}, with an additional scaling for the pathological evolution. We detail the exact regional values in Table~\ref{table:atrophy_values} in the appendix (see \ref{subsec:appendix:synthetic}).

Structural MRIs of 40 healthy subjects from the ADNI database are taken as input to the simulations. For every subject, deformations are simulated for both pathological and healthy settings. The deformation extrapolated 5 times is then applied to the original image in order to simulate 5 years long evolutions.
We then have two matched populations of 40 pairs of images.

\subsubsection{Model estimation}

\begin{figure}[htbp!]
  \begin{center}
    \subfloat[Healthy evolution]{
      \includegraphics[width=0.5\textwidth]{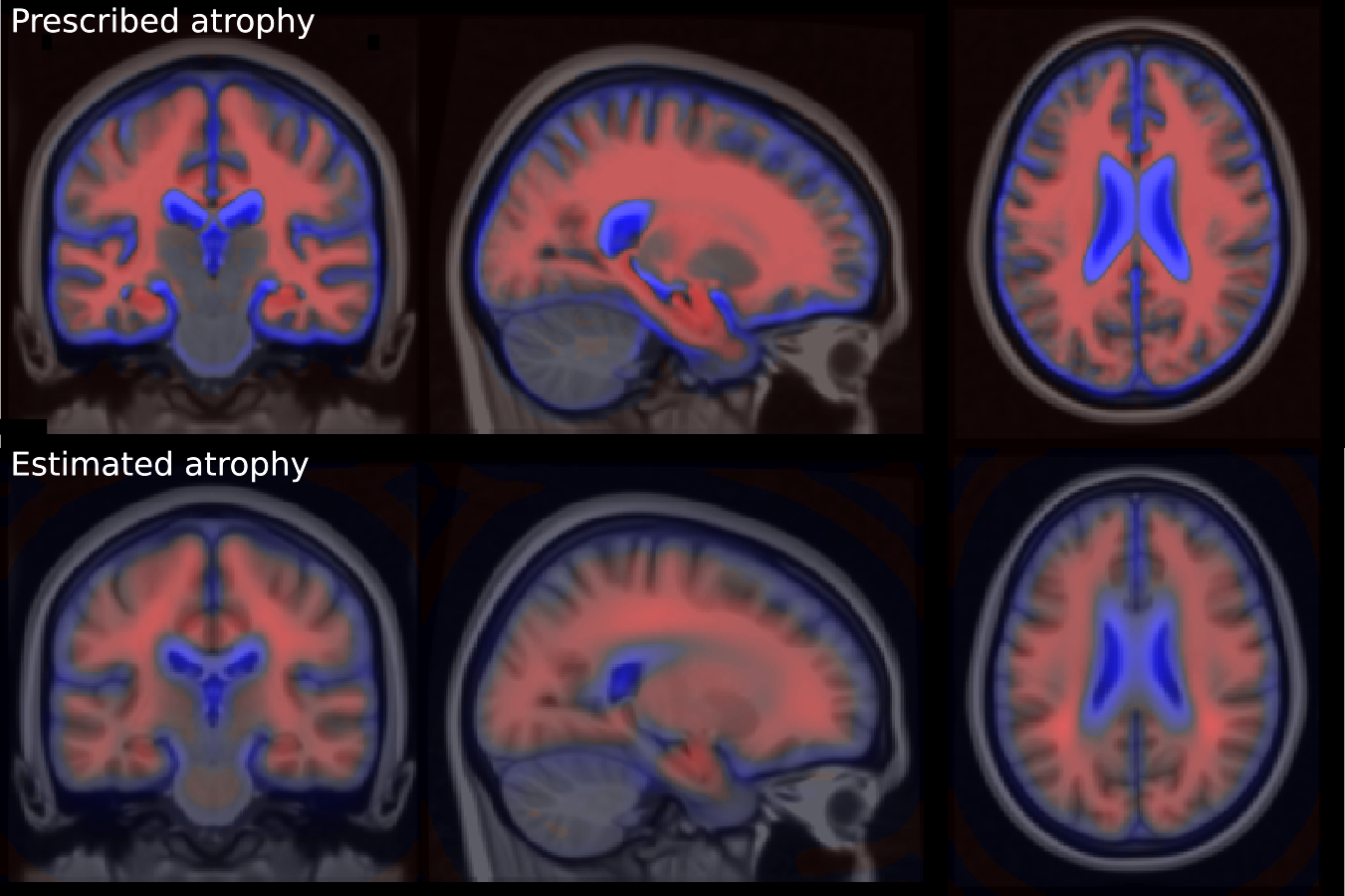}}
        \subfloat[Pathological evolution]{
      \includegraphics[width=0.5\textwidth]{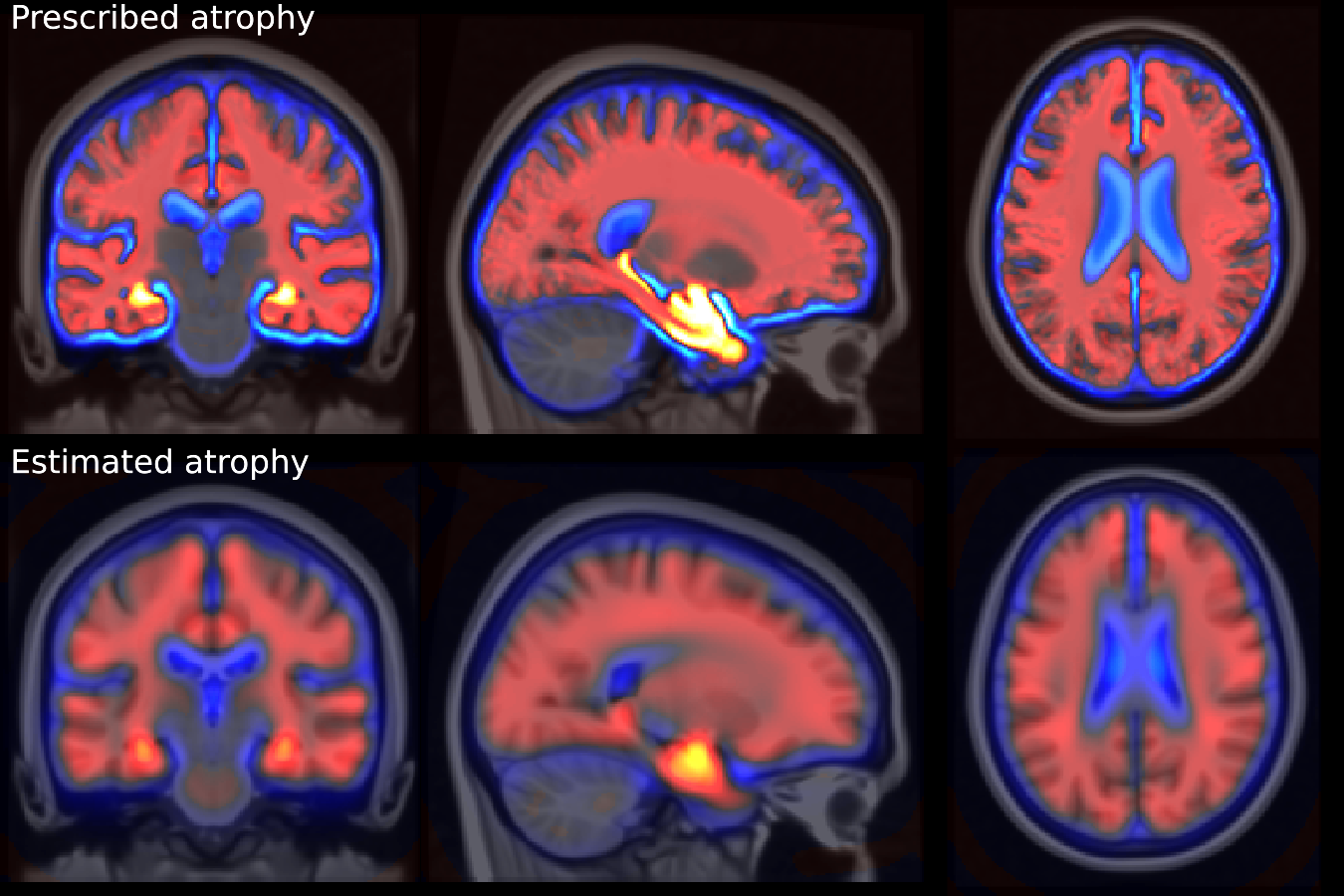}}
    \qquad
            \captionsetup[subfigure]{labelformat=empty}
    \subfloat[relative volume change ($\text{year}^{-1}$)]{
      \includegraphics[width=0.5\textwidth]{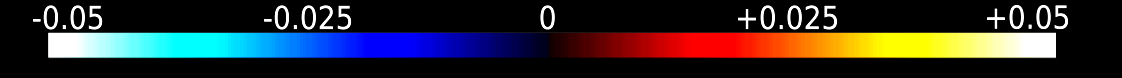}
    }
        \caption{Simulated and estimated atrophy patterns in the template space. Top row: mean in the template space of the prescribed atrophy maps. Bottom row: atrophy maps estimated from the simulated images. Left: healthy simulations. Right: pathological simulations. The estimations are smoother but qualitatively similar to simulated maps.}
    \label{fig:synth_atrophy_map}
  \end{center}
\end{figure}

Individual longitudinal deformations are computed using registration, and the reference anatomy and the template trajectories are built using our framework. The divergence fields associated with these template trajectories can be compared to the prescribed atrophy. Figure~\ref{fig:synth_atrophy_map} shows the average atrophy maps in the estimated template anatomy.

The estimated atrophy patterns are smoothed versions of the simulated ground truth. This effect was already observed~\citep{khanal_biophysical_2016}. First of all, the registration algorithm is unaware of the underlying simulation model and is unable to localize precisely the atrophy in homogeneous areas. Moreover, the spatial regularization of the registration and the parametrisation using SVFs also contribute to the smoothness of the estimated atrophy patterns. This is particularly visible in small (hippocampus) or thin regions (cortex). We therefore have a consistent bias when the atrophy measurements are integrated over the regions (see Figure~\ref{fig:synth_atrophy_boxplot} in appendix). Indeed, the local atrophy is affected by neighboring regions evolving in the opposite direction (the ventricles or the CSF for example). However, we can see that the ratio between the pathological and healthy cases is conserved in every region. It was already noted that quantitative estimation using registration can be biased but can be more reproducible than the segmentation based approaches~\citep{hadj-hamou_beyond_2016, cash_assessing_2015}.

\subsubsection{Imaging biomarkers estimation}

\begin{figure}[htbp!]
  \begin{center}
    \subfloat[Cross-sectional markers at t=5]{
      \includegraphics[width=0.4\textwidth]{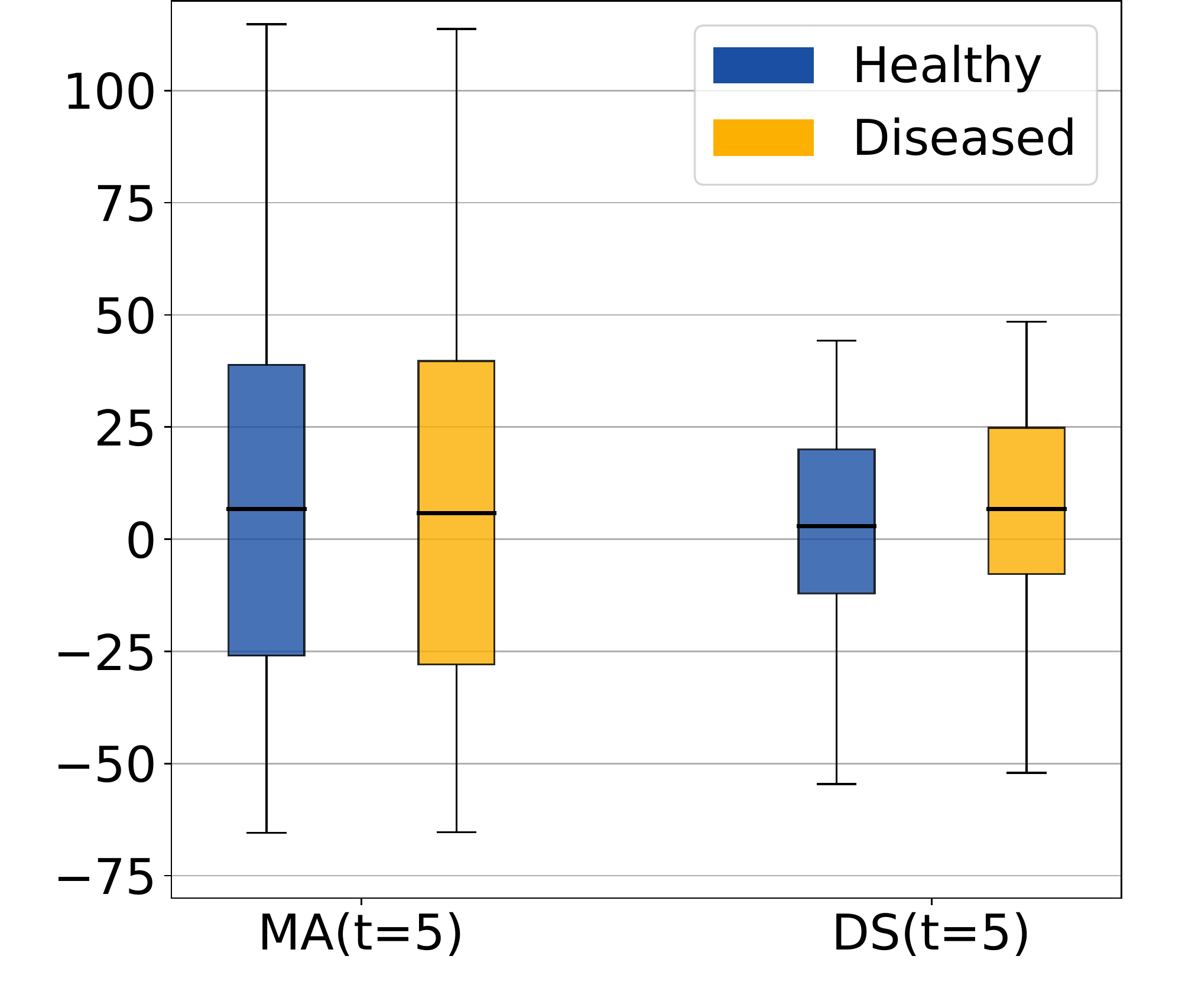}
      \label{fig:synth_bmks_c}}
    \qquad
    \subfloat[Longitudinal evolution]{
      \includegraphics[width=0.4\textwidth]{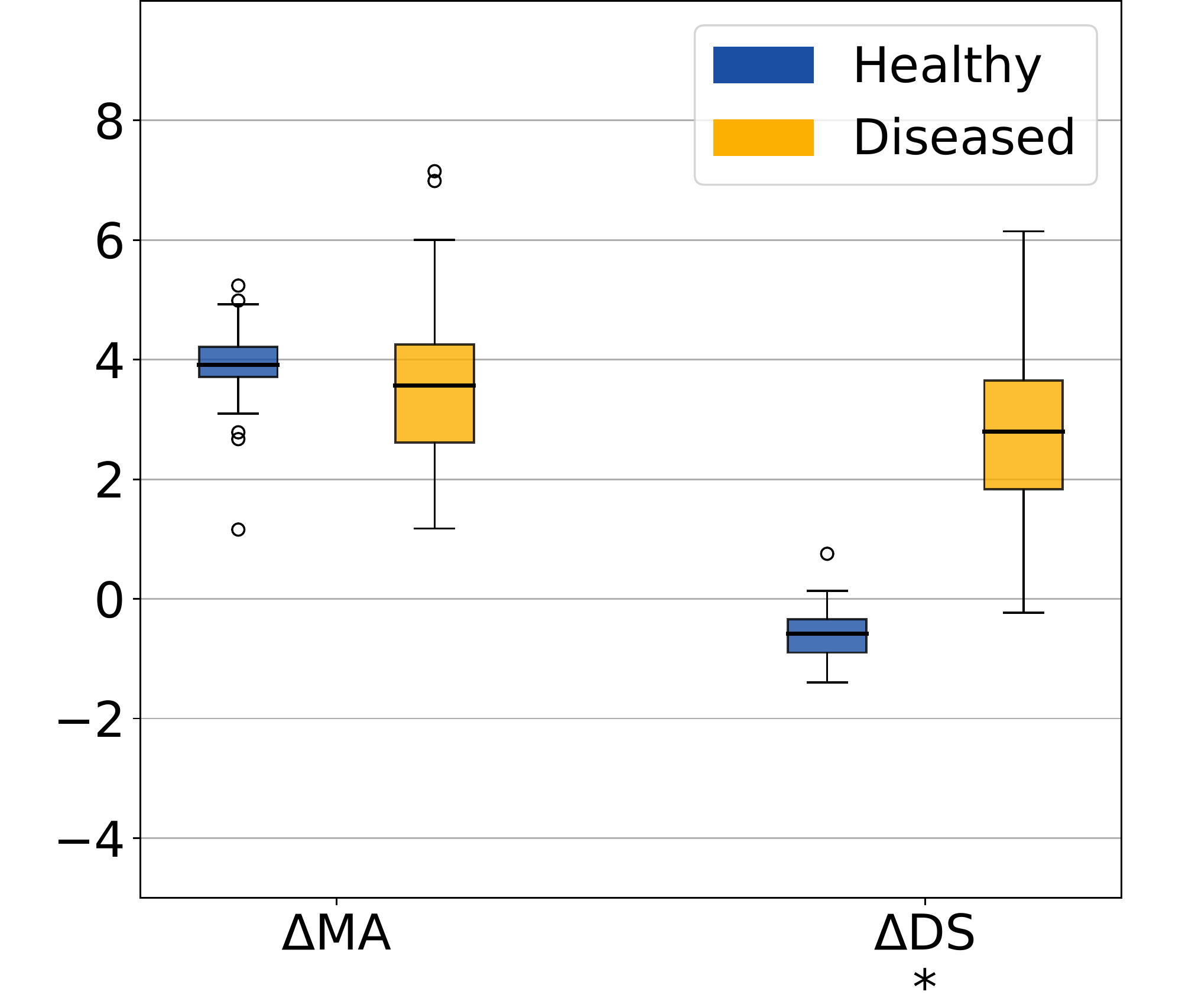}
      \label{fig:synth_bmks_l}}
     \caption{Evolution of the imaging biomarkers estimated on simulated data. MA=morphological age, DS=disease score. The longitudinal evolutions $\Delta$MA and $\Delta$DS are the differences between the two cross-sectional assessments i.e. $\Delta$MA = MA(5) - MA(0). The star indicates that the difference between the healthy and the diseased subjects is significant (p-value < 0.05) for the unpaired t-test. By construction there is no difference between the two populations at t=0. The changes are generally underestimated but the longitudinal evolutions show the stability of the estimation despite a strong inter-subject variability. }
  \label{fig:synth_bmks}
  \end{center}
\end{figure}

The morphological age and the disease score were computed for each image. By construction there is no difference at t=0 (exactly the same images in the two groups). We compare the simulated differences at t=5 and the evolution of the cross-sectional assessments for each subject (see Figure~\ref{fig:synth_bmks_c}).

In this experiment, the initial anatomical variability is important, indeed the standard deviation at t=0 is equal to 48 years for MA(0) and 24 years for DS(0). At t=5, a difference is visible for the disease score but it is still diluted in the inter-subject variability.  It is also possible to extrapolate the evolution to determine how many years of evolution are needed in order to get a significant difference between the healthy and the disease groups. For a significance level of 0.05, the disease score would be significantly discriminant after 13 years of evolution. This figure for synthetic data highlights the slow time-pace of the disease and the interest in modeling and extrapolating the evolutions.

Looking at the evolution of these cross-sectional biomarkers in Figure~\ref{fig:synth_bmks_l}), we see that the measures are relatively stable despite the large inter-subject variability and that the difference measured between two different time points gives a good estimate of the longitudinal evolution. 
In practice the estimations are slightly biased, for example the increase of morphological age $\Delta MA = MA(5) - MA(0)$ is expected to be equal to 5 for both populations while the the mean of the estimation is equal to 3.92 for the healthy group and to 3.57 for the patients group. More importantly, the standard deviation is small in comparison to the standard deviation of the cross-sectional measurement ($\sigma_{\Delta MA}=1.13$ while $\sigma_{MA(0)}=47.8$). For the change of disease score $\Delta DS$, the mean is equal to $-0.6$ for the healthy group and $2.8$ for the patients group and it approximates the ideal expected values (respectively 0 and 5). The variance is also very small with respect to the cross-sectional one. In particular the difference between healthy and diseased subjects is clearly observed for the longitudinal evolution: the difference between the means is equal to 3.2 standard deviation for $\Delta DS$.
To summarize, the cross-sectional measurement gives a relatively stable assessment of the aging and disease progression and the markers' evolution is strongly associated with the clinical diagnosis. Our generative model is able to explain most of the independently simulated changes.

\subsection{Experiments on ADNI data}

Longitudinal T1 sequences were obtained from the Alzheimer's Disease National Initiative (ADNI) database. The ADNI was launched in 
2003 with the primary goal to test whether magnetic resonance imaging (MRI), positron emission tomography (PET), other biological markers, and clinical and neuropsychological assessment could be combined to measure the progression of mild cognitive impairment and early Alzheimer’s disease\footnote{see \url{www.adni-info.org} for more information.}. Subjects are classified according to the evolution of their cognitive diagnosis. Three diagnoses are possible at each time point: normal, mild cognitive impairment (MCI) and Alzheimer's disease. The subjects are also sub-classified according to the positivity of the beta-amyloid 1-42. 
We then have 6 distinct sub-groups: CN- (cognitively normal with negative $A\beta$), CN+ (cognitively normal with positive $A\beta$), MCIs- (MCI stable during the study time-window with negative $A\beta$), MCIs+ (MCI stable with positive $A\beta$), MCIc (MCI converter to AD) and AD (diagnosed with Alzheimer's disease starting from the beginning). The table~\ref{table:adni_demographics} sums up the demographic description of the population.

\begin{table}[ht!]
  \centering
\begin{tabular}{|l|llllll|}
  \hline
  group                     & CN-        & CN+        &  MCIs-      & MCIs+      &  MCIc       & AD         \\
  \hline
  Number of subjects        &        108 &         69 &          96 &        120 &         228 &        203 \\
  Age at baseline           & 73.4 (5.6) & 74.5 (6.5) &  71.1 (7.7) & 73.5 (6.6) &  73.8 (7.1) & 74.5 (7.7) \\
  Gender (female)           &     47.2\% &     56.5\% &      47.9\% &     37.5\% &      41.7\% &     48.3\% \\
  Education (years)         & 16.4 (2.6) & 16.2 (2.7) &  16.2 (2.8) & 16.4 (2.7) &  16.0 (2.9) & 15.0 (2.9) \\
  ADAS13 at baseline        &  9.0 (4.0) &  8.6 (5.0) &  12.0 (4.9) & 14.0 (5.4) &  19.9 (6.7) & 31.4 (7.3) \\
  \hline
\end{tabular}
\caption{Socio-demographic and clinical information of the study cohort. Standard deviations are shown in parentheses. }
\label{table:adni_demographics}

\end{table}

We estimate our template morphology and the template trajectories on a subset of subjects. In order to form this training set, we randomly selected 30 subjects from the CN- group and 30 from the AD group. To reduce the variability associated with the estimation of the model, these subjects were selected among the ones with strictly more than one followup acquisitions.
In the following we distinguish between the \emph{training} set of 30+30 subjects used to build the model and the remaining \emph{testing} set (with in particular 78 CN- subjects and 173 AD subjects).

\subsubsection{Estimation of the normal aging and the disease-specific template trajectories}
\label{subsec:res:trends}

\begin{figure}[ht!]
  
  \begin{center}
    \subfloat[Normal aging template trajectory]{
      \includegraphics[width=0.4\textwidth]{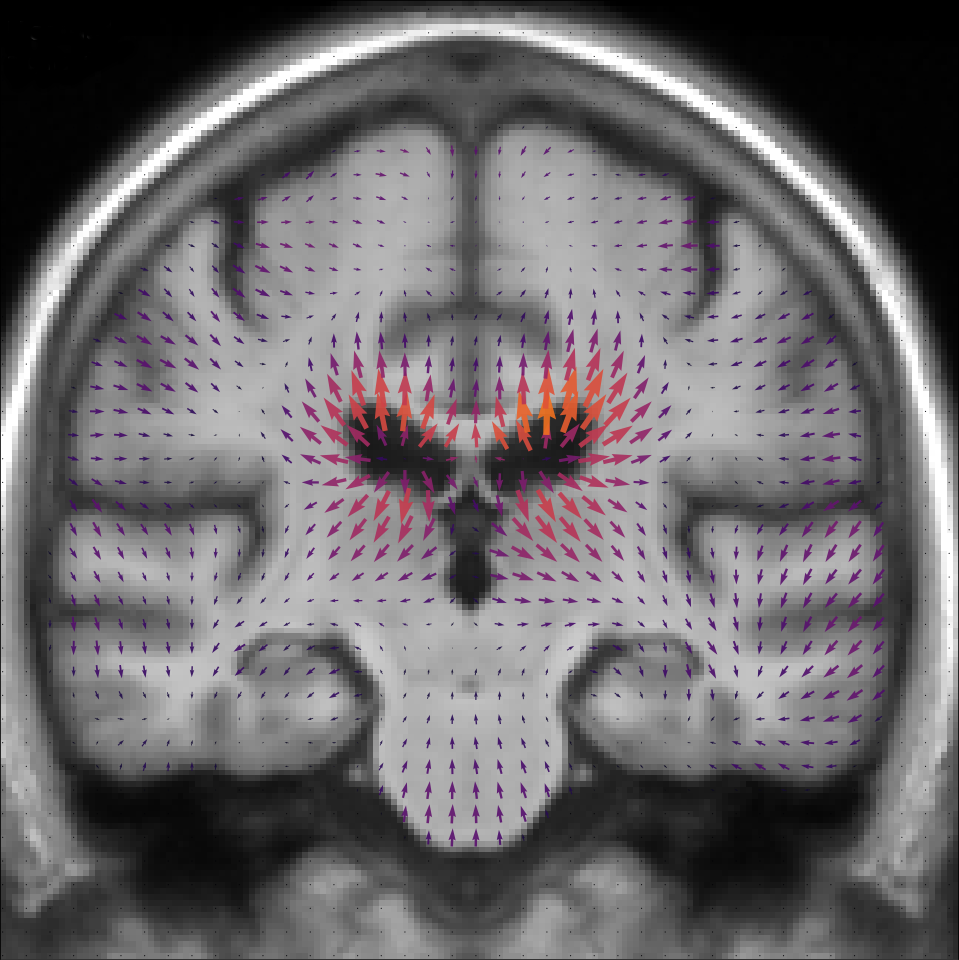}}
    \subfloat[Disease specific template trajectory]{
      \includegraphics[width=0.4\textwidth]{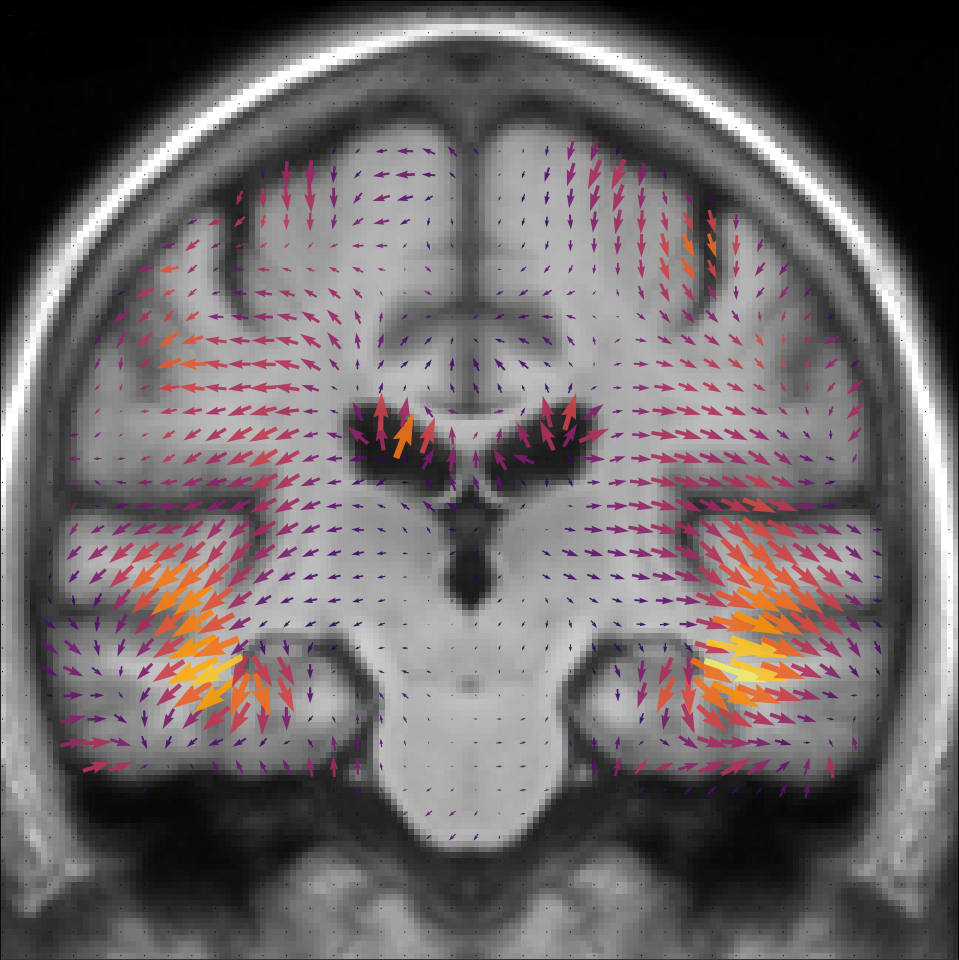}}
    \qquad
            \captionsetup[subfigure]{labelformat=empty}
    \subfloat[($\text{mm} \cdot \text{year}^{-1}$)]{
      \includegraphics[width=0.5\textwidth]{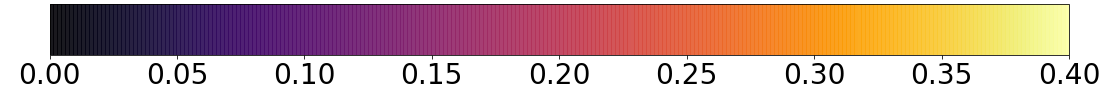}}

    \caption{Template image and SVFs parametrizing the two template trajectories SVFs. Left: normal aging trajectory $\mathrm{v}_{A}$ showing a ventricular expansion related to a global atrophy. Right: disease specific trajectory $\mathrm{v}_{D}$ showing specific patterns, especially in the temporal lobes around the hippocampi areas. The color encodes the amplitude of the velocity at each position.}
  \label{fig:atlas}
  \end{center}
\end{figure}

The template anatomy is an average of the healthy subjects anatomies, so its age corresponds to the mean group age $\lambda_0 = 73.46$ y.o. The result of the estimation is shown in Figure~\ref{fig:atlas}. The estimated normal aging template trajectory is characterized mainly by ventricular expansion caused by the atrophy of the surrounding regions. Disease specific changes are widespread in the brain with a strong emphasis on the temporal areas.

\begin{figure}[htbp!]
  \begin{center}
    \subfloat[Atrophy along the normal aging template trajectory]{
      \includegraphics[width=0.8\textwidth]{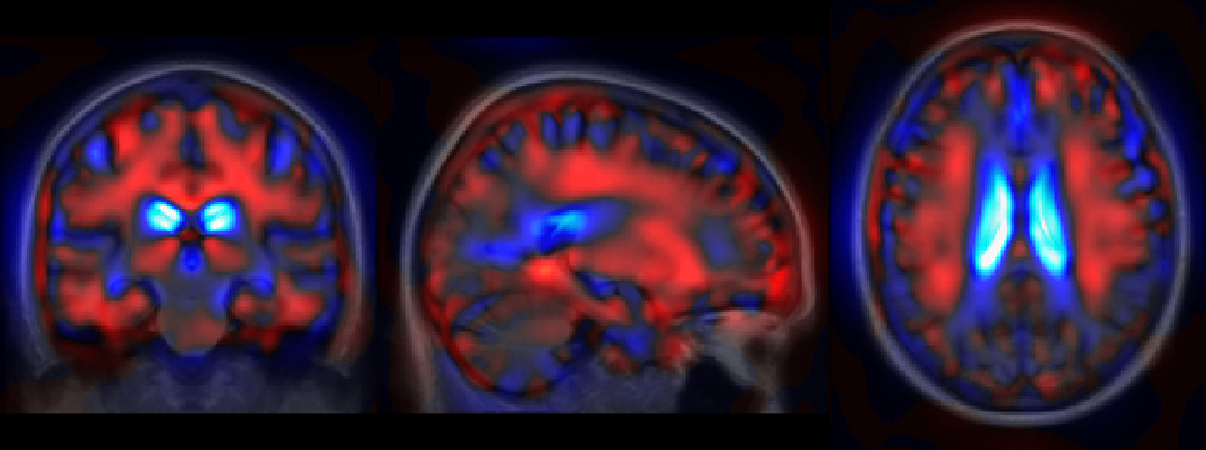}}
    \qquad
    \subfloat[Atrophy along the disease specific template trajectory]{
      \includegraphics[width=0.8\textwidth]{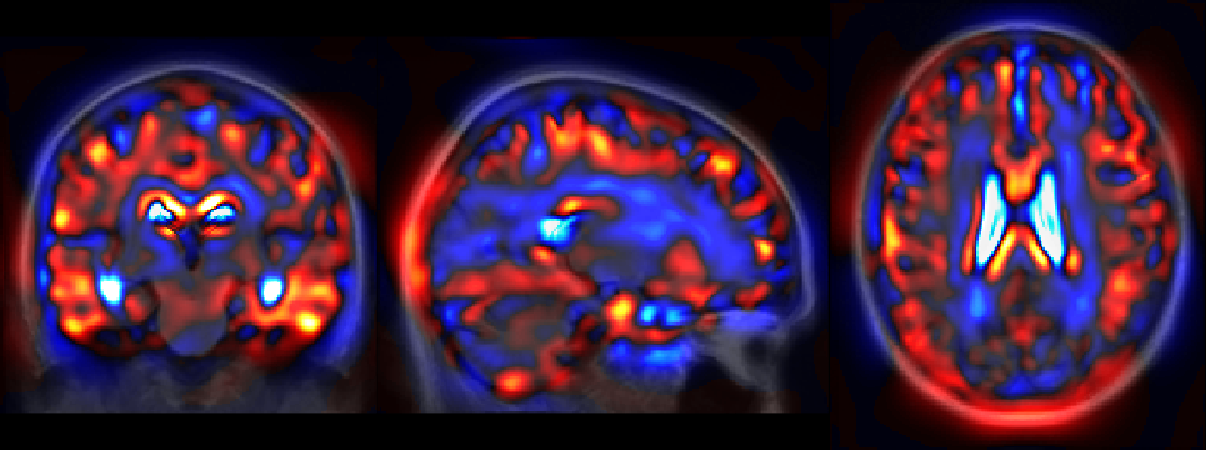}}
    \qquad
            \captionsetup[subfigure]{labelformat=empty}
    \subfloat[relative volume change ($\text{year}^{-1}$)]{
      \includegraphics[width=0.5\textwidth]{fig_div_colorbar.png}}
    
  \caption{Atrophy measured by the divergence of the SVFs parametrizing the two template trajectories.}
  \label{fig:atlas_logjac}
  \end{center}
\end{figure}

The local atrophy can be measured by the divergence of the velocity field. The atrophy associated with each template trajectory is shown in Figure~\ref{fig:atlas_logjac}. For the normal aging, we see a well spread and mild atrophy pattern in the whole brain. The disease specific atrophy is particularly strong in the temporal area and is mainly located near the cortical surface. The precise localization of the atrophy is always difficult with a morphometrical approach but the atrophy patterns are similar to what was already observed in the past for healthy subjects and AD patients~\citep{hadj-hamou_beyond_2016}.

These morphological evolutions can be compared to the normal aging model and the mean residual for the AD subjects in~\citet{lorenzi_disentangling_2015}. The deformation characteristics and the magnitude of the atrophy are really similar. Our anatomy looks sharper which may be partially explained by the use of the template plane in the estimation. More importantly, differently from the work of~\citet{lorenzi_disentangling_2015}, the joint estimation of a disease model in addition to the normal aging one provides us with a direct comparison between both processeses.

\begin{figure}[htbp!]

  \begin{center}
                    \includegraphics[width=1.0\textwidth]{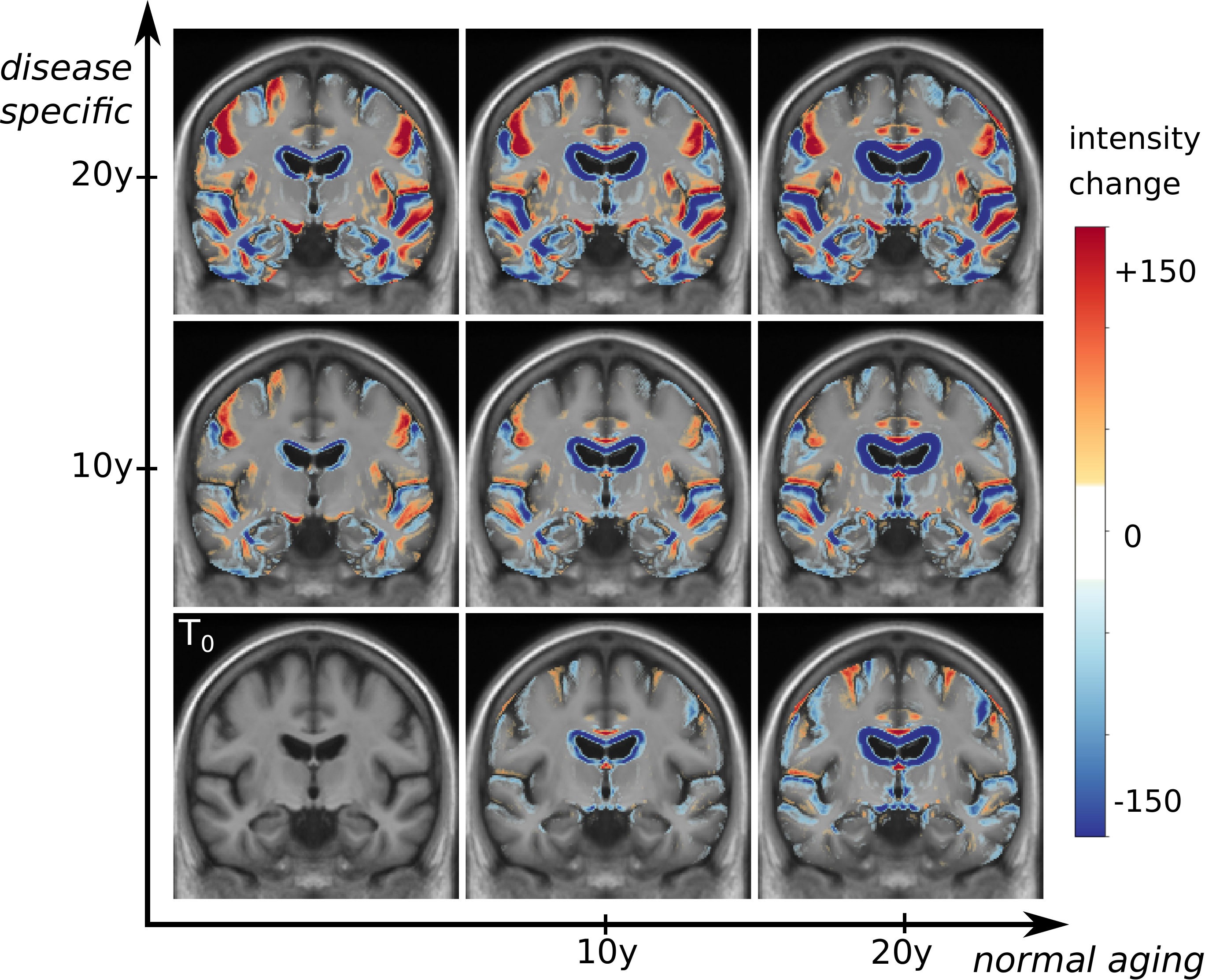}
        \caption{Representation of the 2D parametric template subspace generated using the template morphology $T_0$ and the two template trajectories $v_{A}$ (horizontally) and $v_{D}$ (vertically). In this figure, the bottom row correspond to a healthy evolution, and the diagonal (from bottom left to top right) to a typical pathological evolution. We also represent the voxel-wise intensity differences between the images and $T_0$ to highlight the boundary shifts between tissues and CSF.}
    \label{fig:atlas_plane}
  \end{center}
\end{figure}

The generative model can also be used to directly visualize the modeled morphologies. In particular the reference template plane $\mathcal{T}$ described in Figure~\ref{fig:schema_model} can be sampled to shows the evolution of the template morphology in the two main directions. In Figure~\ref{fig:atlas_plane}, we choose to represent the evolution over 20 years in both directions which is comparable to the longitudinal span of our data-set. Indeed the IQR of baseline age in the training set is equal to 9.3 years (and the total span is 29.9 years wide). The overlaid difference of intensity is used to show the changes at tissue boundaries. The global atrophy and the expansion of the ventricles is clearly visible for the aging evolution. The pathological changes are associated with smaller structures but the shrinking of the hippocampi, the atrophy of the temporal lobes and also the widening of the sulci (related to the cortical thickening) are visible.

\subsubsection{Intra-subject variability of our progression markers}
\label{subsec:res:validation-intra}

In \citet{lorenzi_disentangling_2015}, a morphological age similar to our measure was shown to be correlated with the chronological age and also that advanced AD stages were associated with ``morphologically older'' brains.
To go further, we want to show that our proposed model represents also the aging at the individual level. For multiple acquisitions of a same subject, an aging measurement is expected to increase smoothly with time. If the subject is healthy, we can expect a linear increase with a slope of 1. We would also like to see an increase of disease score for the patients, while for the healthy subjects this marker should be stable and close to 0. Therefore, the disease score is expected to specifically characterize the pathology (high and increasing disease score). The morphological age is expected to be more associated with the global evolution of the population over time and to be independent of the clinical condition.

\begin{figure}[ht!]
  \centering
  \includegraphics[width=1.0\textwidth]{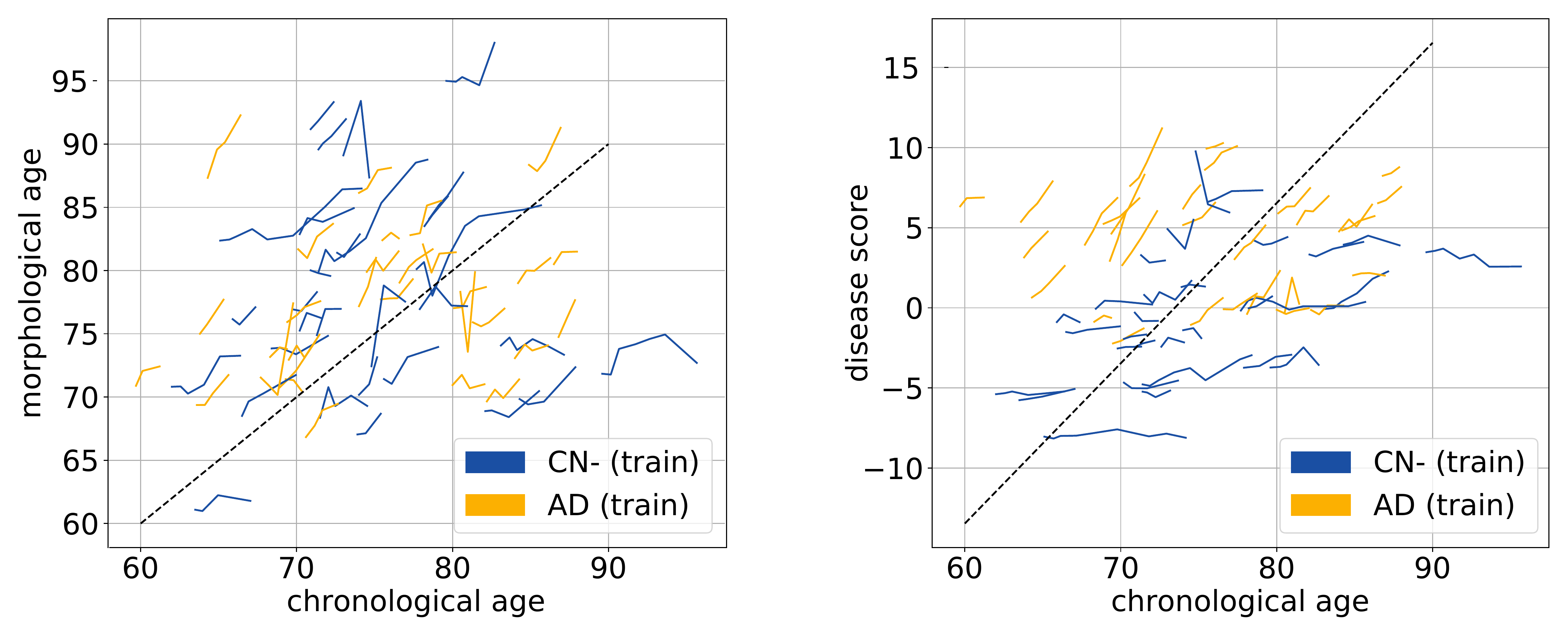}
    \caption{Evolution of cross-sectional markers for every subject of the two training sets. [Left] chronological age, the dashed line corresponds to the expected evolution of healthy subjects i.e. the morphological age is equal to the chronological age. [Right] disease score, the dashed line is the expected pathological evolution.}
  \label{fig:spaghetti}
\end{figure}

The morphological age and the disease score are computed for each subject at each time point. Figure~\ref{fig:spaghetti} shows the evolution of these cross-sectional measurements. First, at the population level, subjects are generally associated with a morphological age similar to their chronological age even though its variability is quite high. Second, for each subject, the evolution is mostly linear and the morphological age steadily increases. Third, the disease score is almost constant for healthy subjects and steadily increases for the AD subjects. Finally, we note that the AD subjects look older, age faster and, most of all, have a higher disease score than the healthy ones.

A linear random effects model can help us to quantify these observations. The model is fitted, for both morphological age and disease score, with fixed effects on age and sex and a random intercept and slope for each subject. The focus is set on the analysis of the regression for the CN and the AD groups. For each coefficient of the regression, the confidence intervals are given for one standard deviation of the estimation.

The model is first fitted to the morphological age measures in the CN group leading to a coefficient of $0.26 \pm 0.11$ for the fixed effect of age while the mean subject slope is $0.10 \pm 0.02$. Both are significantly positive. In comparison to the same model without the random slope the relative improvement brought by the intra-subject linear evolution is significant by a large margin (p-value inferior to $10^{-6}$ for the likelihood ratio test). The regression has also a positive (but not significantly) coefficient for male subjects ($1.81 \pm 1.2$) meaning that male morphologies looks older (similar to a $7$ years shift). Concerning the disease score, we also observe a relatively good fit of the linear model. The evolution is generally slower with $0.12 \pm 0.1$ for the fixed effect of age and $0.12 \pm 0.01$ for the mean individual slope.

For the AD group, the linear model is also well adapted (p-values inferior to $10^{-6}$ for the likelihood ratio test). The main remark is probably that the intra-subject slopes are in average more important than for healthy subjects (around $0.52 \pm 0.06$ for the morphological age and $0.71 \pm 0.05$ for the disease score) while the fixed effect related to age of $0.23 \pm 0.07$ (for the morphological age) and $0.14 \pm 0.04$ (for the disease score) are more similar to the one observed previously.

\subsubsection{Cross-sectional discriminating power}
\label{subsec:res:xsectional}

We want to study the relation between the observed disease progression and the proposed markers. We start with a discriminant analysis using only the first image available for each subject. 
\begin{figure}[ht!]
  \centering
  \includegraphics[width=1.0\textwidth]{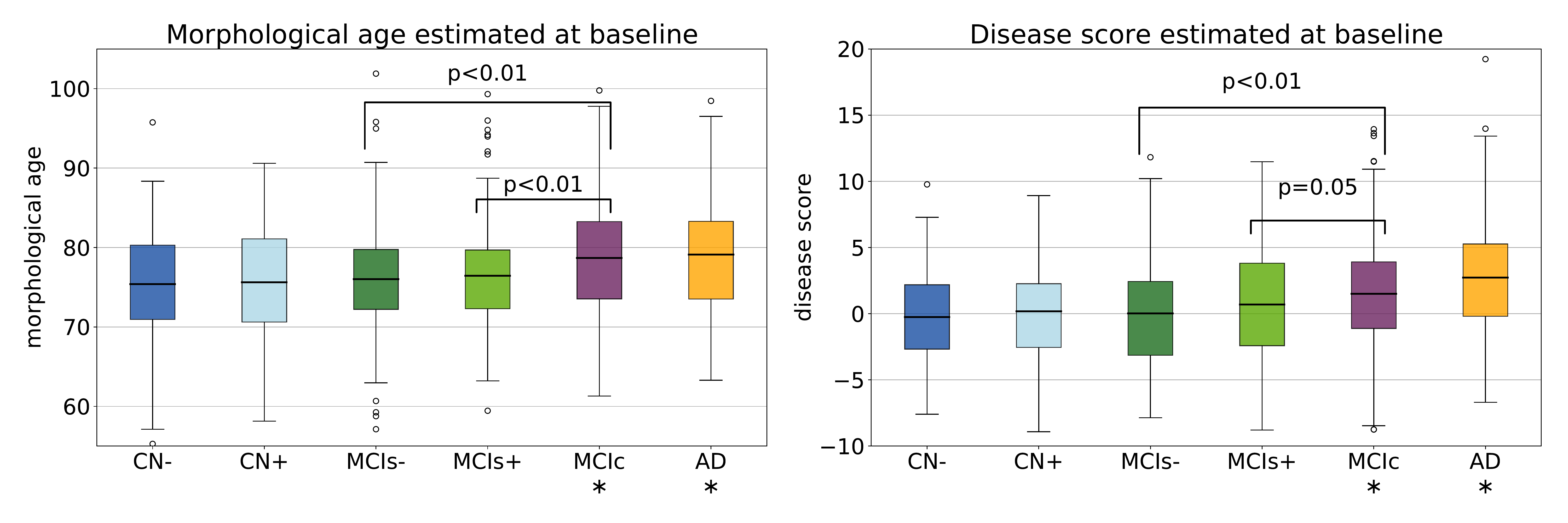}
  \caption{Box-plot of the group-wise markers estimated at baseline for the clinical groups. Stars below the name of the group indicate a significant difference to the CN- group for a t-test at the level 0.05. Both markers gradually increase towards more advanced disease states.}
  \label{fig:violins_cross}
\end{figure}

Figure~\ref{fig:violins_cross} shows the distribution of the estimations for each group. These results are related to global differences in brain shape observable cross-sectionally between clinical groups. We see a gradual increase of both markers towards more advanced disease states. Significant differences in morphological age and in disease score are observed between the control group CN-(train) and both the MCIc and AD groups.  Moreover, the difference between the MCI stable and the MCI converters is stronger for the morphological age while the disease score better differentiates MCIs- and MCIs+. As such, the morphological age is more associated with the general cognitive degradation while the disease score seems more correlated with more AD-specific biomarkers.

We also perform a simple linear classification task between the MCIs and MCIc groups using this two cross-sectional markers. A SVM linear classifier is fitted to the full data-set to perform the binary classification\footnote{The classification task was performed using the SVC module of the scikit-learn python library~\citep{scikit-learn}}. The error penalty weights are adjusted between the two classes to balance the trade-off between false positive and false negative rate. The mean classification accuracy using a 10-folds cross-validation scheme is equal to $0.59$. The linear decision function is equivalent to the projection on the SVF $\mathrm{v}_A - 0.003 \mathrm{v}_D$, so the differences between MCIs and MCIc subjects is, in our model, only associated with the aging trajectory while the disease specific changes do not seem to have an impact before the conversion. For comparison, for the same experiment between CN and AD, the linear classifier corresponds to the projection on $\mathrm{v}_A + 0.49 \mathrm{v}_D$ so approximately $(v_{bn} + v_{ad}) / 2$, i.e. the mean evolution of the whole population. Of course, in both cases we do not reach the performance of state-of-the-art dedicated algorithm but it allows us to see how both markers are associated with the diagnosis. Moreover this discriminant approach could be extended by using information in a subset of targeted areas.

\subsubsection{Regional analysis of the progression}
\label{subsec:res:population-trajectories}

\begin{figure}[ht!]
  \centering
  \includegraphics[width=1.0\textwidth]{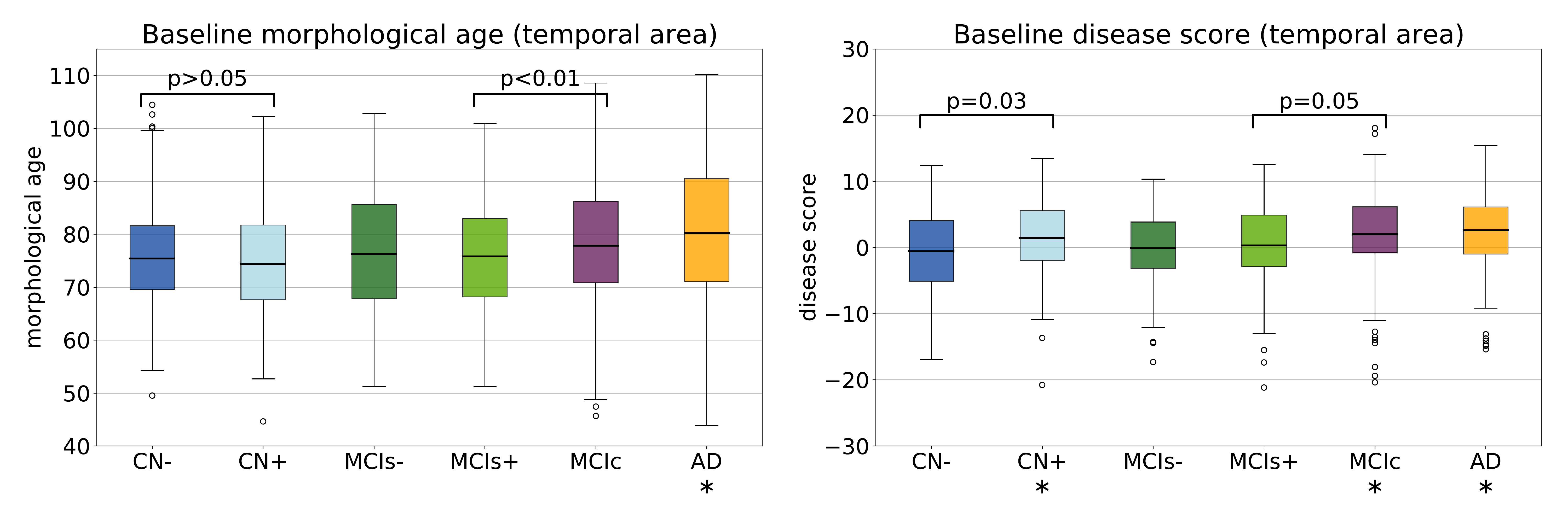}
    \caption{Box-plot of the markers estimated in the temporal lobes. Stars below the name of the group indicate a significant difference to the CN- reference for a t-test at the level 0.05. The area known to be related to the AD makes the disease score estimation less sensitive to the overall noise.}
    \label{fig:temporal_markers}
\end{figure}

In the context of Alzheimer's disease, specific morphological changes are known to be non-uniformly distributed. This spatial information can be taken into account in our model using a regional segmentation. In this section we focus on the temporal area which is often associated with Alzheimer's disease~\citep{double_topography_1996}. Using the AAL atlas~\citep{tzourio_automated_2002}, we segment the temporal lobe of our template anatomy. The mask is then used to compute the regional morphological age and disease score for each subject. These markers only encodes the morphological differences in this specific area visible at the time of acquisition (i.e. at baseline in this case). Results are shown in Figure~\ref{fig:temporal_markers}.

The region is more adapted to the disease score than to the morphological age model. Indeed, for a healthy subject the deformations in this area are really small. However the choice of a disease-adapted region is improving the performance of the disease score. It is now able to capture early specific changes and the difference between CN- and CN+ is significant.

\subsubsection{Longitudinal evolution of the markers}
\label{subsec:res:longitudinal}

\begin{figure}[ht!]
  \centering
      \includegraphics[width=1.0\textwidth]{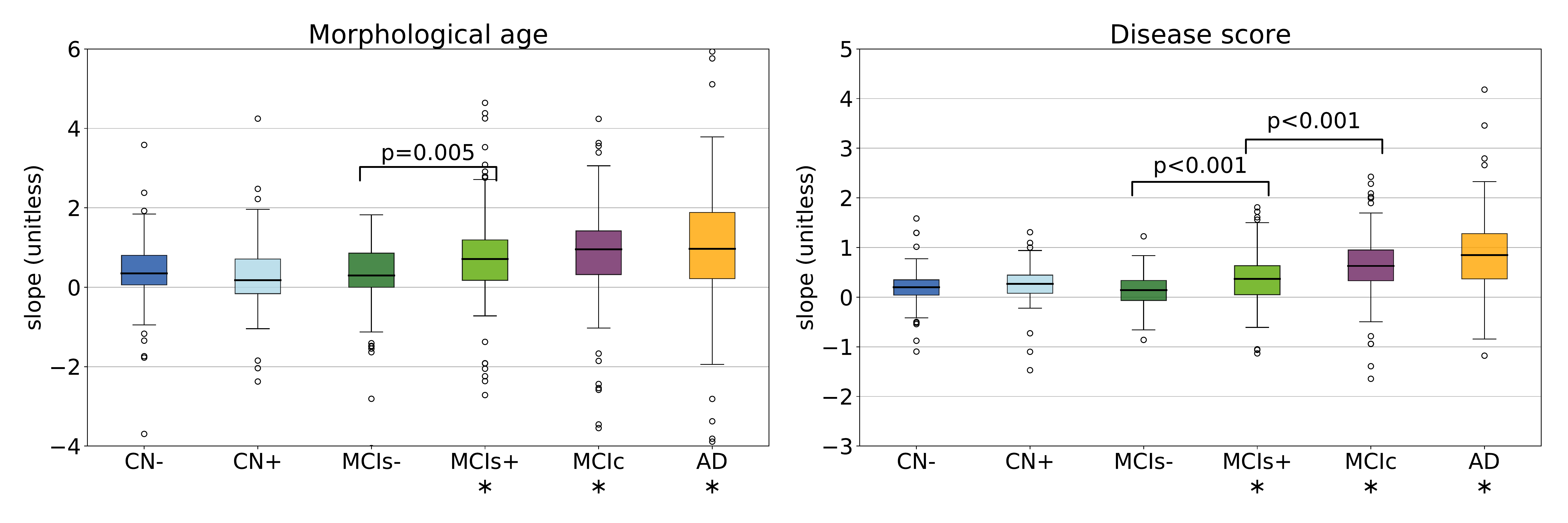}
  \caption{Box-plot of the rate of evolution of the markers computed using individual linear regressions. Stars indicate a significant difference to the CN- reference for a t-test at the level 0.05. Top row shows the results for the whole brain while the bottom row shows the result for the temporal lobe only. A gradation is visible from the CN to the AD subjects.}
  \label{fig:violins_slopes}
\end{figure}

To explore more in details the longitudinal evolution of these markers, a linear model is fitted to the individual evolutions.
The intercept can be interesting as it aggregates the measure at every time point and helps reduce the noise but more importantly the slope can be very informative. Results are shown in Figure~\ref{fig:violins_slopes} for the whole brain markers.

A progressive evolution, from CN- to AD, is visible for the morphological age with subjects evolving faster and faster.
Concerning the disease score, the evolution is almost negligible for CN- and MCIs- and relatively slow for CN+ and MCIs+ while the changes are clearly visible for MCIc and AD.
Significant differences are visible between healthy subjects (CN-) and MCIc or AD subjects or even between MCI stable and MCI converters, but also between MCIs- and MCIs+ (or more generally between subjects with negative amyloid or positive amyloid marker).
This may indicate that our measures are able to capture the global progression of the disease. The changes are larger for diagnosed patients but similar patterns of evolution are observed in the early stages of the disease.
A significant difference is also observable between the CN- and CN+ group for the temporal disease score slope.

The disease score evolution is close to zero for the healthy subject and close to one for the pathological one but more generally, the slopes are in average smaller than their expected values. For example the average disease score evolution in the AD group is equal to $0.82$ and this discrepancy is particularly important for the morphological age slope of the CN- group that is only equal to $0.33$ (instead of $1$). This bias was already observed previously and can be in part explained by the estimation procedure (see ~\ref{subsec:appendix:bias}).

\subsection{Generating diagnosis driven morphological evolution}
\label{subsec:res:generating-trajectories}

One of the main advantages of our model is its ability to be generative. From a pair of biomarkers (morphological age, disease score) we can generate a corresponding morphology and to deform a specific subject anatomy in the directions defined by the template trajectories. In this section, the model is used to generate plausible morphological evolutions of a subject, for several diagnosis condition, and compare them to the observed one.

\subsubsection{Modeling the group evolution}

\begin{figure}[ht!]
  \centering
  \includegraphics[width=0.48\textwidth]{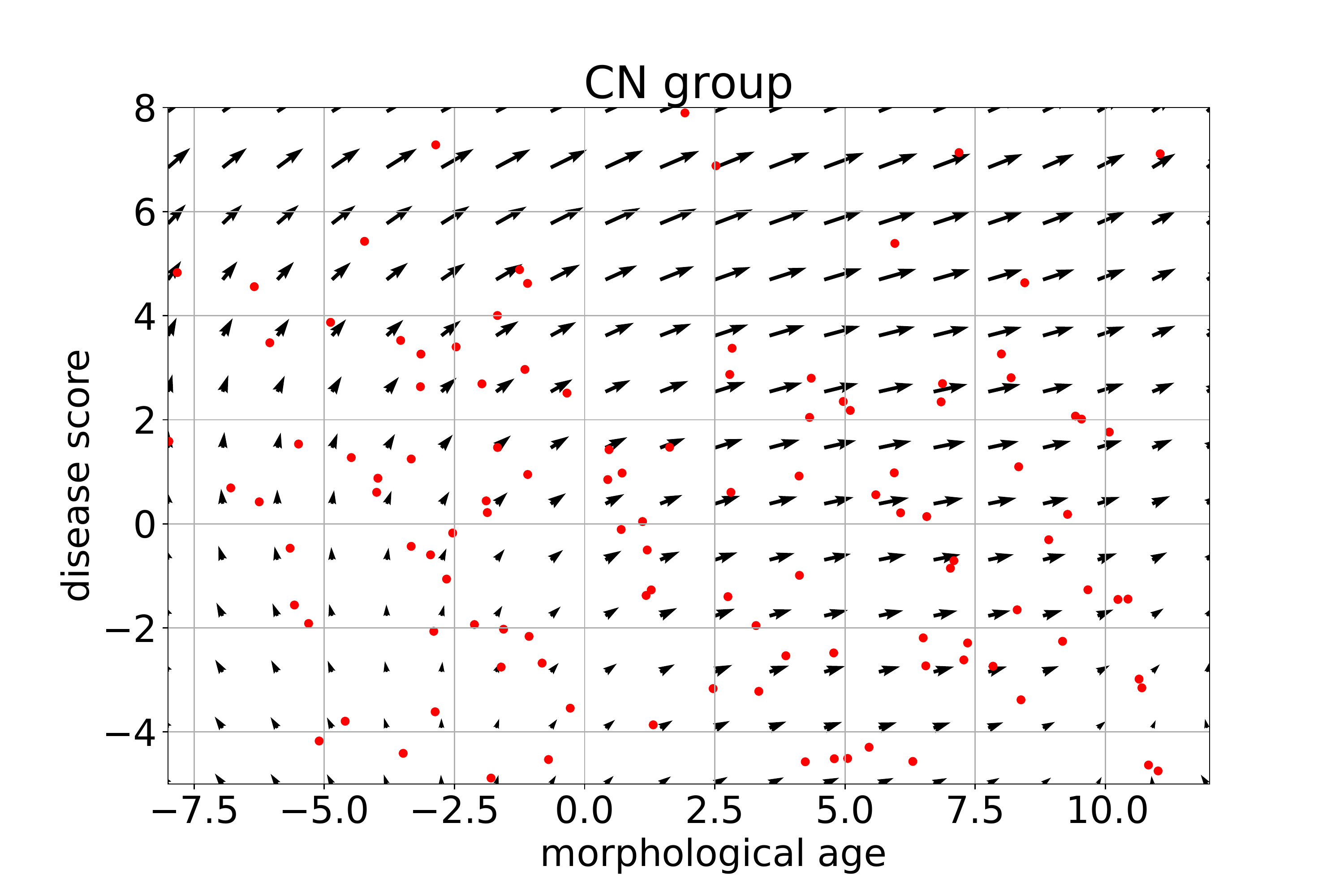}
  \includegraphics[width=0.48\textwidth]{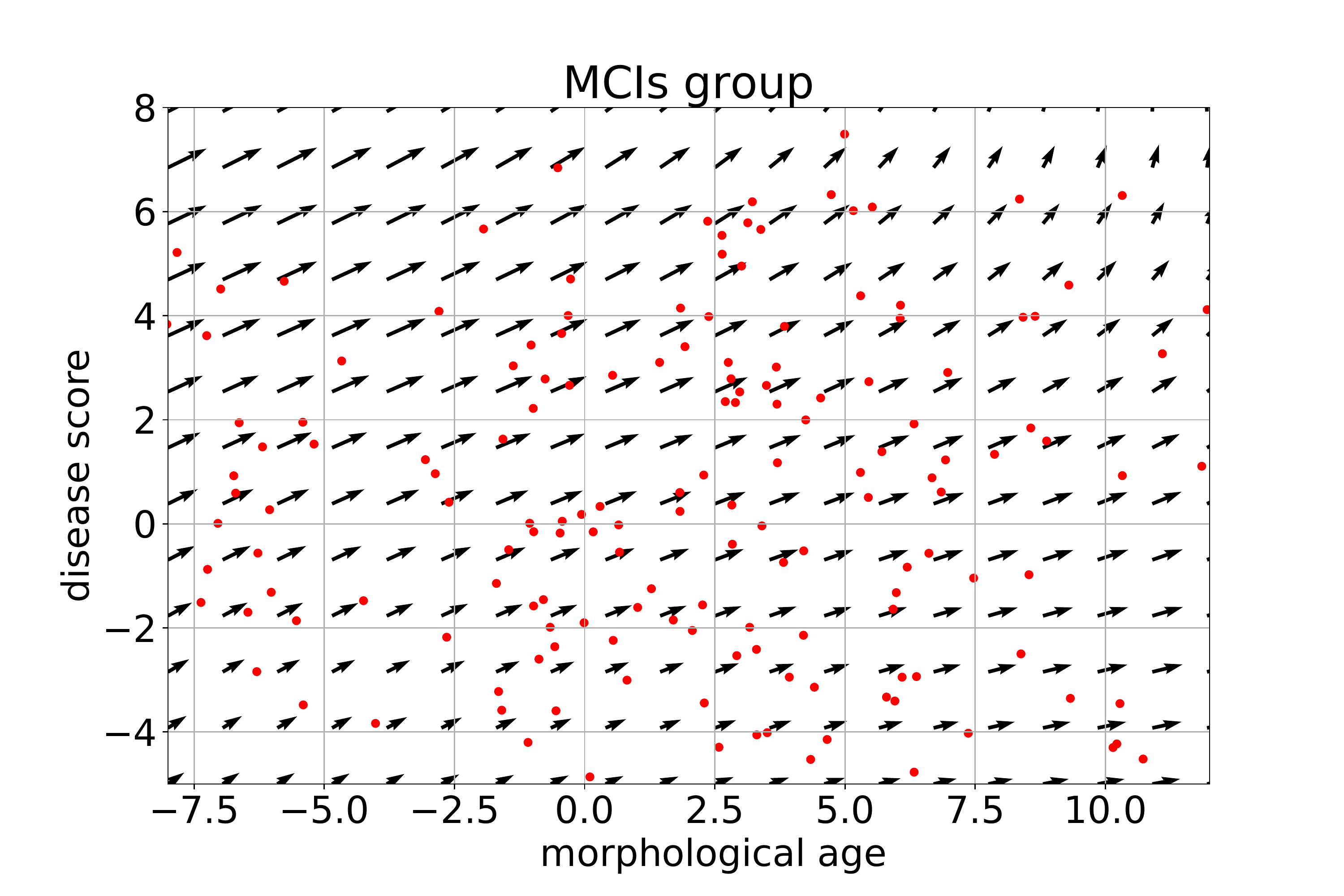}
  \includegraphics[width=0.48\textwidth]{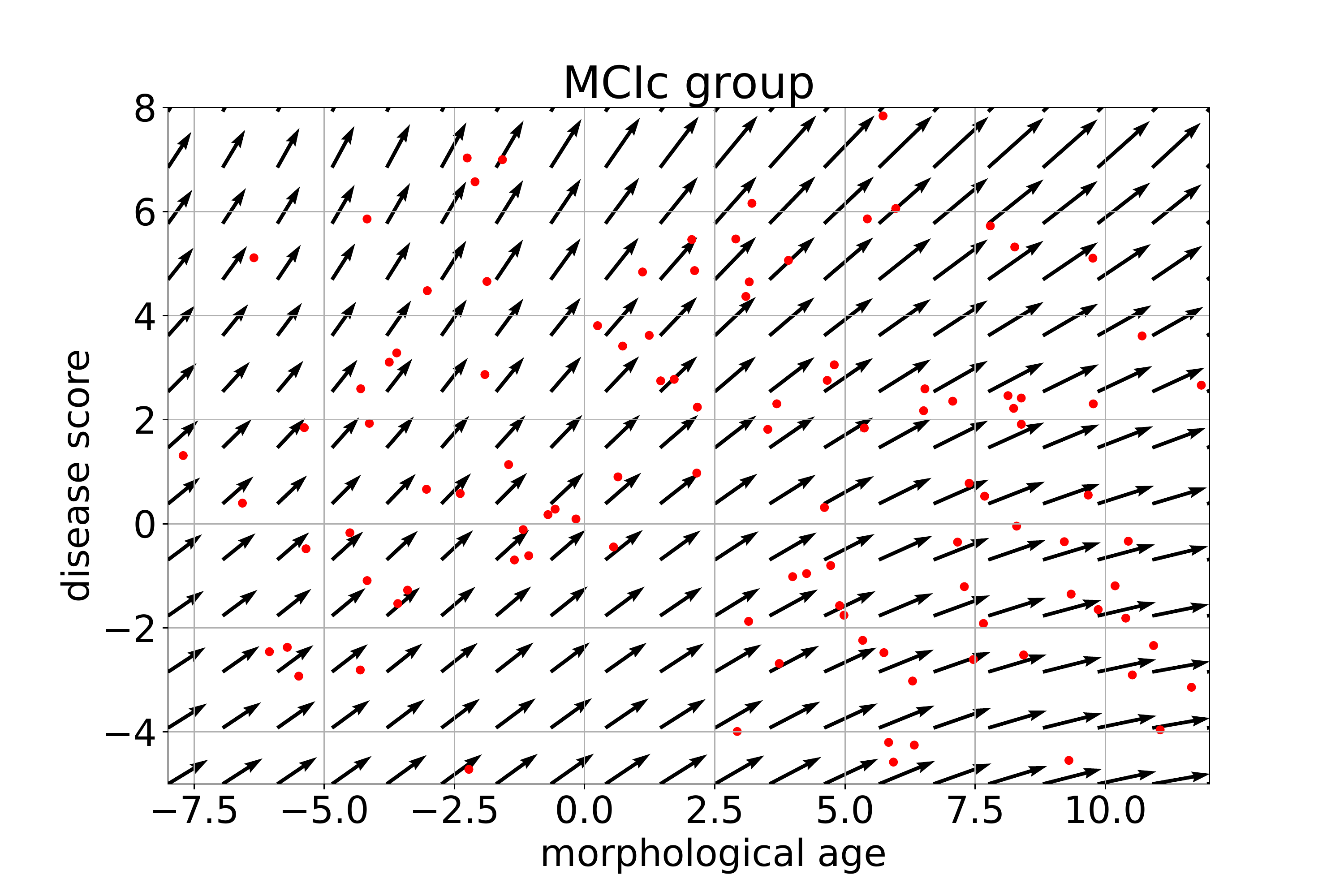}
  \includegraphics[width=0.48\textwidth]{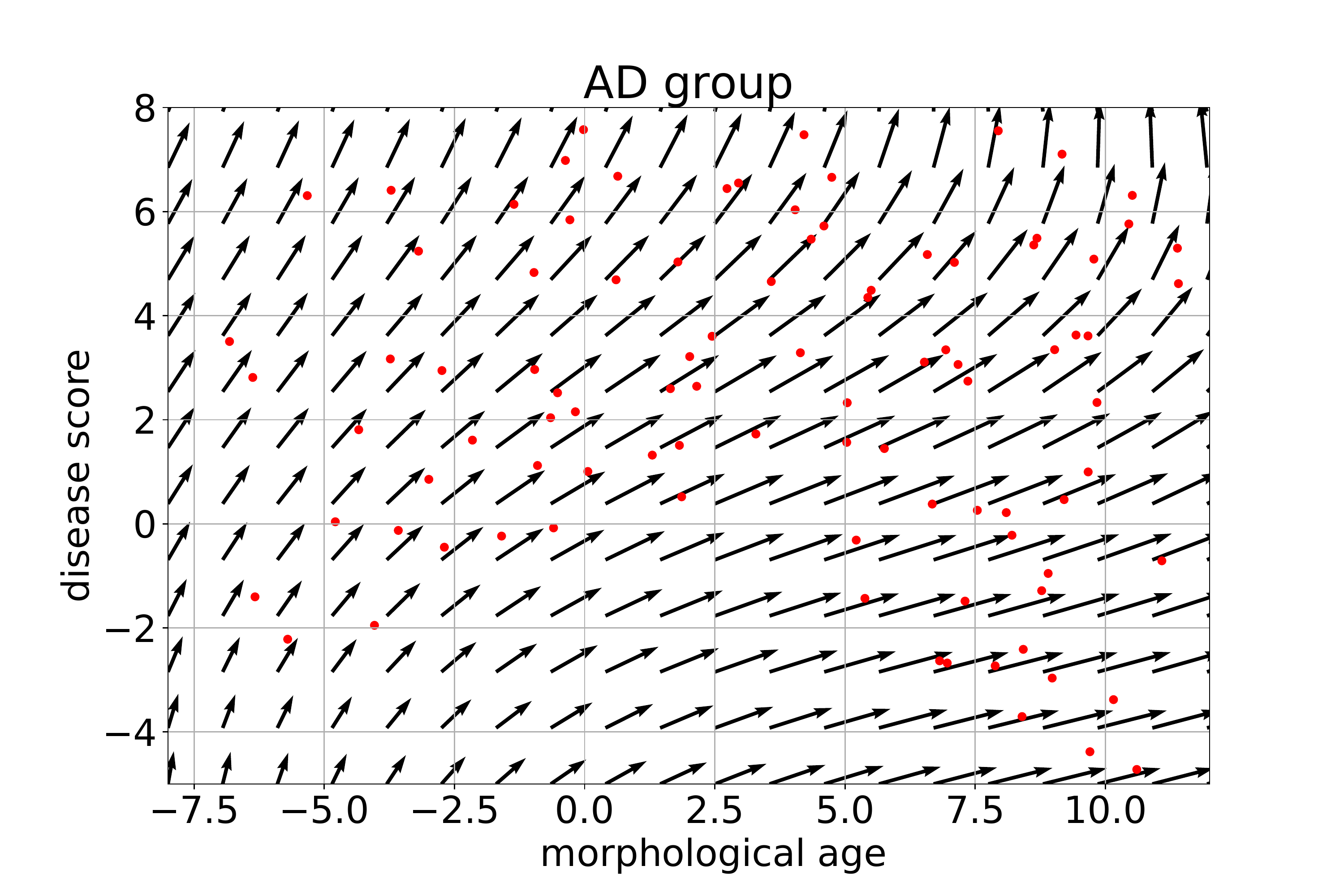}
  \caption{Results of the kernel ridge regression for the markers' evolution for the CN, MCIs, MCIc and AD groups. It shows the regressed vector field with data points shown in red. Amplitude and orientation variability is visible between the groups (with stronger and more vertical evolution for AD than for CN) but also within each group giving non-linear morphological evolutions.}
  \label{fig:streamlines}
\end{figure}

In what follows, each subject is associated with a pair (position, speed) in the marker plane using the coefficient of the linear regression. For each group, a continuous vector field is regressed from these observations. This field estimates the local average speed at every point. In order to keep a local consistency in the markers space, we use a kernel ridge regression with an RBF kernel. The spatial scale is set to 10 years, for both the morphological age and the disease score axis, to get large scale patterns despite the high inter-subject variability. The regularization weight, which does not seem to have a large effect on the result, is set to 1.
Results are presented in Figure~\ref{fig:streamlines} for the CN (i.e CN-, CN+ together), MCIs, MCIc and AD groups. The figure is centered on the high data density domain (as expected the extrapolation can be less reliable in lower density sectors).

Differences in amplitude, i.e. speed of evolution, and orientation are clearly visible between the groups and are in agreement with the linear regression results shown in the previous section. In particular we see a progressive amplitude increase from the CN group to the AD group.

These diagrams also help to describe the variability within the same clinical group.
For the CN group we can distinguish between the low disease score and low morphological age area (in the bottom left) where in average the changes are negligible, and the rest where there is a slow horizontal evolution. This pattern suggests that the healthier and younger subjects are morphologically stable and do not show the same visible aging process.
The MCI stable evolution is relatively uniform and in average with slightly larger amplitudes but overall similar to the CN one.

The MCI converters however show a stronger and more vertical evolution. We should also note that subjects with high morphological age and low disease score (bottom right) seem to follow a different, more horizontal evolution implying a fast morphological aging but less important disease specific changes.
The AD group confirms this trend and in fact MCIc and AD look very similar. The mean evolution is strong and more vertical. A main evolution is visible from bottom left to top right with a sightly more horizontal part in the middle giving this global tangent-like aspect. Beside, a horizontal evolution, similar to what was observed for the MCIc model, is also visible in the bottom right. This difference of evolution suggests a possible stratification of the disease in two sub-categories.

\subsubsection{Generating a subject specific evolution}

The regression model can then be used to simulate the evolution of a given subject. We choose here to model the evolution of MCIc subject in order to predict the changes subsequently observed around the time of diagnosis. Several evolutions were computed for the subject 0361. This subject is chosen among the MCIc subjects because it has the longest time interval, here 8.5 years, between the first and the last acquisition. From the starting point, the markers at t=0, we integrate a trajectory using the speed given by the regression model at each point. Each clinical group is associated with its own model and then a different trajectory is computed for each diagnosis.

\begin{figure}
  \centering
  \subfloat[Subject 0361]{
    \includegraphics[width=\textwidth]{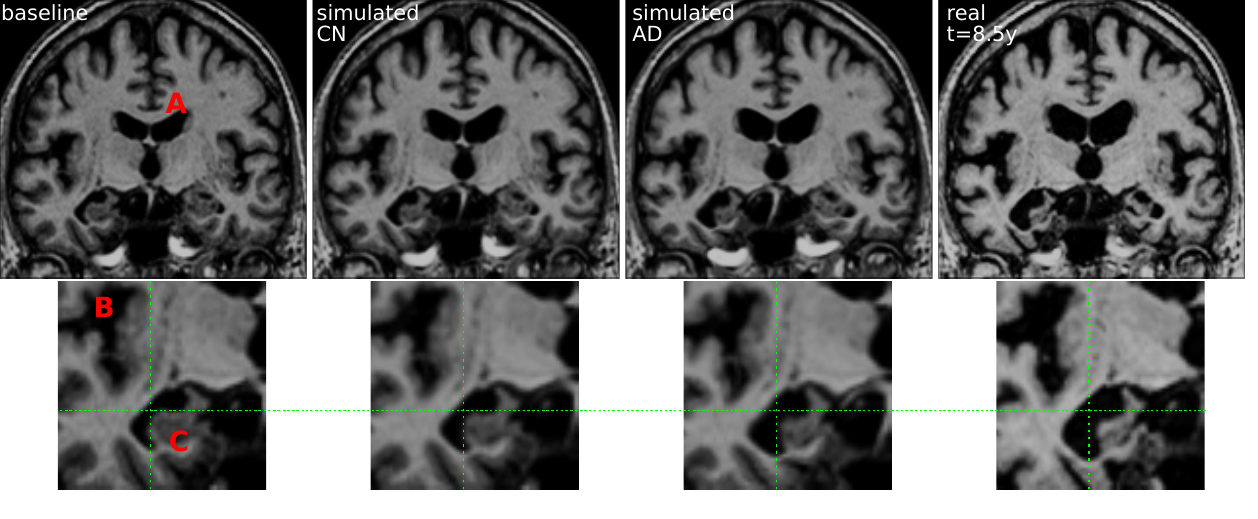}}
  \qquad
  \subfloat[Subject 0566]{
    \includegraphics[width=\textwidth]{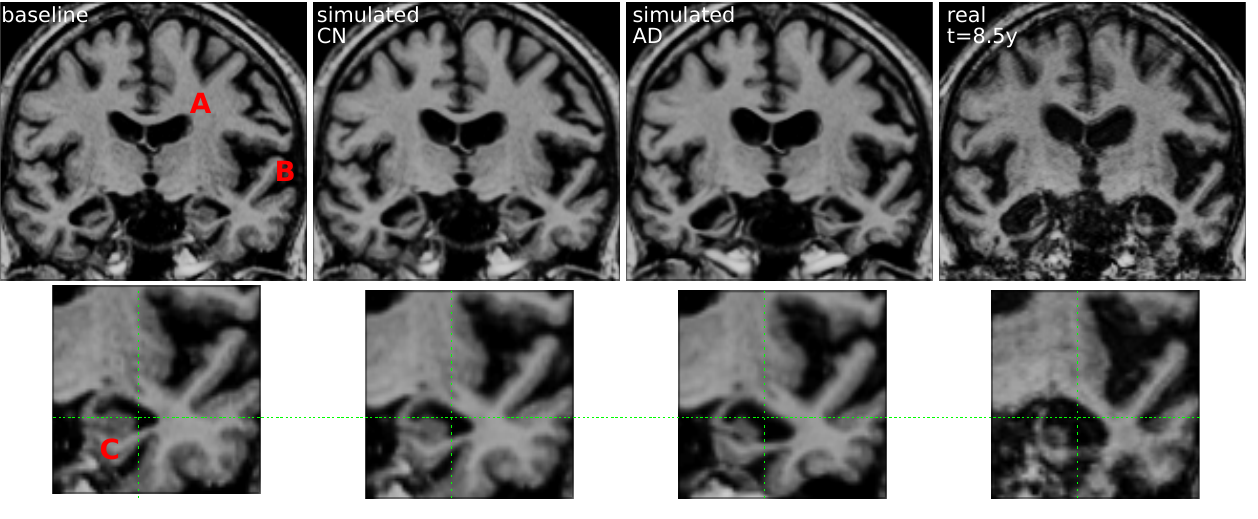}}
    \caption{Evolution of the two subjects over 8.5 years. Both subjects were diagnosed MCI (for 5y and 3y) then AD. From left to right: (1) real image at baseline, (2) simulated image from the baseline image using the healthy (CN) evolution, (3) using the pathological (AD) evolution, (4) real image at t=8.5y. The second row zooms in the most interesting areas between the ventricle (A), the lateral sulcus (B) and the hippocampus area (C). In both cases, even if the simulated changes do not match the full extend of the real case, the atrophy is visible in the sulcus and the hippocampus and there is a difference in shape and size of the ventricles.}
  \label{fig:evo0361}
\end{figure}

These markers changes can be directly translated in brain images to visualize the morphological evolution. Figure~\ref{fig:evo0361} shows the results for the end point of the trajectories for the CN and AD models. Images are generated by deforming the baseline image using the simulated deformation transported in the subject space. A bias correction is applied to the markers' estimation using the value estimated on the synthetic dataset (the measured changes only correspond to 80\% of the expected value).

For both subjects, the cortical atrophy, in particular in the temporal lobe, is clearly visible in the real image and to a lesser extent in the AD simulation. The expansion of the ventricles is always clearly visible, but even in the AD simulation, the volume change is inferior to the observation. Subject 0566 is, from the baseline, generally more atrophied and by comparison it makes some real changes less visible (in particular the sulci widening).
This difference in baseline anatomy is interesting here for two reasons: first the deformations match correctly the anatomical structures (the hippocampi for example) despite the anatomical variability, second it generated two different evolutions for the same diagnosis. For the AD simulation, the deformation seems stronger in the temporal lobe and relatively weaker in the ventricles for the second case (0556).
These evolutions are learned from population trends and even if they are not predictive for a particular subject, they are not aberrant in comparison with the real evolutions.
  Overall, the morphological changes simulated looks realistic even if they do not match perfectly the observed changes. Other aspects of the evolution are hard to quantify and often poorly documented. For example the evolution of the shape of the ventricles of subject 0361 is different in the three images. It may be related to different spatial distribution of the degeneration in the brain and inherent mechanical constraints. We also observe a global motion towards the bottom of the temporal lobe and a local rotation in the image for the real case and the AD simulation.

\section{Discussion and perspectives}
\label{sec:discussion}

In this work, we proposed a novel deformation-based approach to measure the progression of normal and pathological processes from their effects on brain morphology. In the context of Alzheimer's disease, it provides a simple description of the brain morphological evolution for elderly patients using only two degrees of freedom: an aging measurement and a disease score. The advantages come from three main properties.

First, we disentangle the aging and the disease progression using interpretable image-based biomarkers.
Second, these markers are cross-sectional assessments and are consistent for intra-subject longitudinal analyses. They can be seen as alternative aging measurements compatible with ongoing biological processes. In particular the disease specific evolution appears to be associated with a positive amyloid marker even in prodromal stages.
Third, we show that the markers and the generative model can be used in a personalized image simulation setting. It allows us to generate smooth and realistic evolutions for several diagnosis conditions.

Biological (or here morphological) age estimates were proven to be interesting to analyze the patient condition; here, in addition, the disease score is used to get a simple marker of the disease progression. The joint modeling gives a more complete description of the disease progression than a brain-age metric. The evolution is not seen as a simple accelerated aging process or the divergence from the normal evolution. On the contrary, it seems possible to capture the general worsening with the morphological age while the disease score measures an additional evolution that is more specific to the disease. Both patterns of evolution appeared to be related with the development of Alzheimer's disease and this approach provides an intuitive interpretation and a simple decomposition of the morphological changes observed in mass-univariate morphometric approaches. Further analysis should be conducted to analyze the relation of these complementary patterns to clinical and cognitive variables.

Moreover the ideas behind the decomposition and the longitudinal/cross-sectional estimation of the parameters could be used in a different setting: the geodesic description is much more general than the SVF framework applied to structural MRIs. Other geodesic parametrizations could be more practical in another setting and even for the same image data, an LDDMM approach, for example, would generally concentrate the deformation more on the high intensity gradient interfaces while the SVFs model more spatially diffuse deformations.

Several approaches have been developed to model the longitudinal shape evolution using geodesic modeling~\citep{singh_hierarchical_2013}. Here we suggest that decoupling the estimation of individual trends using parallel transport could allow a more accurate and non-linear description of the disease while keeping a similar hierarchical mixed-effect structure.

As could be expected, a single variable is not enough to precisely describe the morphological changes that can occur while aging. Likewise, the inter-subject variability limits the accuracy of the modeling of the evolution of individual subjects. Some limitations come from the error and the approximation in the estimation of the model, others are related to aspects that are not taken into account and would require to modify the approach.

\subsection{Approximations in the model estimation}

One of the limitations of our approach is the difficulty to accurately estimate the markers. For example, some time points look like outliers when we perform the estimation for successive images of the same subject. This might be due to MR distortions that cannot be completely corrected by the registration. As a result, some artifacts are still visible in the estimated deformations. A better understanding of the effects of these distortions on the registration results would be useful to improve our additive deformation noise model. A first step could be to work with data where the ground-truth deformation is more controlled (scan-rescan images for example).

In this work, we used an orthogonal projection using the $L_2$ scalar product in the SVF space to define the subject specific deformation. This choice is arbitrary and using another region of interest could change the result. A possible alternative would be to match the subject with the closest morphology in the model for a metric more adapted to deformations. We could also try to decorrelate the two information, using ICA for example.

We saw, using synthetic and longitudinal data, that we were under-estimating the evolution speed.
This bias is partially explained by the estimation procedure that does not take into account the uncertainty on the population parameters of the model (see~\ref{subsec:appendix:bias}). This problem cannot be easily solved for two reasons: first the bias caused by the estimation of the template anatomy using the healthy controls groups is hard to quantify, second an unbiased estimation would be far more complex and would not be possible without the full knowledge of the training set for every new subject. 
Another source of error can be the registration algorithm: the inter-subject registrations are larger than the intra-subject ones and no transitivity is guaranteed.

\subsection{Limitations of the description}
\label{subsec:discussion:limitations}

Probably the main limitation of our approach is the inter-subject variability of the markers. The markers are very stable in longitudinal but the anatomical variability is not completely absorbed by the biomarker estimation and they are sensitive to the subject differences. The use of an explicit or even implicit model of the shape of a normal brain would help to makes them more specific to time related changes.

We could also note that the template anatomy is estimated using healthy subjects only. This modeling choice may introduce a bias towards the healthy population. We argue that the explicit modeling of the disease in our model could help mitigating this issue. Moreover, this choice makes the origin of the morphological age and the disease score simpler to define. Finally, pulling all the subjects could also enable a more accurate estimation~\citep{marchewka_influence_2013} but the relatively large and homogeneous populations involved does not make it critical here.

In this context, the use of a single reference anatomy to parametrize the template space could also be discussed. Here, for example, it introduced a bias toward a certain age because of the way we composed deformations. A multi-atlas approach could be a better solution if a single anatomy is not enough to describe our population~\citep{blezek_atlas_2007,sabuncu_image-driven_2009}. And we could do something similar to what was proposed by~\citet{rohe_barycentric_2016} to intricate the registration and the template subspace prior.

Regarding the estimation of the template trajectories using individual longitudinal SVFs transported in the template, potential bias associated with the asymmetrical use of the baseline image (in processing and in modeling) has been highlighted~\citep{reuter_within-subject_2012} and the imbalance in number of time points or total longitudinal span should be taken into account for a more reliable estimation. However, \citet{hadj-hamou_longitudinal_2016} has shown that the asymmetry caused by the non centrality of the time point in the longitudinal sequence is not completely relevant in the stationary velocity field framework because the SVF are expressed in Eulerian coordinate and should be identical at all time points.

\subsection{Perspectives}
\label{subsec:discussion:perspectives}

The use of segmentation to compute the progression markers in multiple regions would be another way to extend the description. We showed that using a segmentation of the temporal lobes could tighten the link between the morphological markers and early clinical conditions. However the question of the regional interactions is not addressed in this work and the spatial analysis of brain deformation remains a research topic.

Our model of the morphological evolution is generic and of low dimensionality. Consequently it only partially captures the changes related to AD and, in practice, the subject specific field $w^k_{r}$ is actually encoding changes that are related to the disease and its variability but are not currently modeled. To enhance the description, the model should integrate the observed and latent clinical condition of the subjects. A description that goes beyond the  global cognitive diagnosis would be interesting to pursue in the temporal description of the morphological changes in order to better describe the disease progression and capture this evolution from a healthy state to a pathological one. 

Beside, the observed longitudinal evolution is also influenced by various factors such as sex, genetic, other pathologies or even the image acquisition protocol. The ability to handles these covariates has shown its utility to better describe the brain morphological evolution~\citep{muralidharan_bayesian_2016}.

These extensions would also improve the generative aspect. Coupling our approach with a proper disease progression model, and using a mixture model for the trajectories, would enable the generation of morphological trajectories in a more diverse setting to explore and sample the range of possible evolutions.

\section*{Acknowledgment}
\label{sec:acknowledgment}

This work has been supported by the Université Côte d'Azur and by the Inria Sophia Antipolis - Méditerranée, ”NEF” computation cluster.

Data collection and sharing for this project was funded by the Alzheimer's Disease Neuroimaging Initiative (ADNI) (National Institutes of Health Grant U01 AG024904) and DOD ADNI (Department of Defense award number W81XWH-12-2-0012). ADNI is funded by the National Institute on Aging, the National Institute of Biomedical Imaging and Bioengineering, and through generous contributions from the following: AbbVie, Alzheimer's Association; Alzheimer's Drug Discovery Foundation; Araclon Biotech; BioClinica, Inc.; Biogen; Bristol-Myers Squibb Company; CereSpir, Inc.; Cogstate; Eisai Inc.; Elan Pharmaceuticals, Inc.; Eli Lilly and Company; EuroImmun; F. Hoffmann-La Roche Ltd and its affiliated company Genentech, Inc.; Fujirebio; GE Healthcare; IXICO Ltd.;Janssen Alzheimer Immunotherapy Research \& Development, LLC.; Johnson \& Johnson Pharmaceutical Research \& Development LLC.; Lumosity; Lundbeck; Merck \& Co., Inc.;Meso Scale Diagnostics, LLC.; NeuroRx Research; Neurotrack Technologies; Novartis Pharmaceuticals Corporation; Pfizer Inc.; Piramal Imaging; Servier; Takeda Pharmaceutical Company; and Transition Therapeutics. The Canadian Institutes of Health Research is providing funds to support ADNI clinical sites in Canada. Private sector contributions are facilitated by the Foundation for the National Institutes of Health (www.fnih.org). The grantee organization is the Northern California Institute for Research and Education, and the study is coordinated by the Alzheimer's Therapeutic Research Institute at the University of Southern California. ADNI data are disseminated by the Laboratory for Neuro Imaging at the University of Southern California.

Authors have no conflict of interest regarding this article

\bibliography{biblio}

\begin{thebibliography}{51}
\providecommand{\natexlab}[1]{#1}
\providecommand{\url}[1]{\texttt{#1}}
\expandafter\ifx\csname urlstyle\endcsname\relax
  \providecommand{\doi}[1]{doi: #1}\else
  \providecommand{\doi}{doi: \begingroup \urlstyle{rm}\Url}\fi

\bibitem[{Alzheimer's~Association} and {others}(2017)]{association_2017_2017}
{Alzheimer's~Association} and {others}.
\newblock 2017 {Alzheimer}'s disease facts and figures.
\newblock \emph{Alzheimer's \& Dementia}, 13\penalty0 (4):\penalty0 325--373,
  2017.
\newblock URL
  \url{http://www.sciencedirect.com/science/article/pii/S1552526017300511}.

\bibitem[Arsigny et~al.(2006)Arsigny, Commowick, Pennec, and
  Ayache]{arsigny_log-euclidean_2006}
V.~Arsigny, O.~Commowick, X.~Pennec, and N.~Ayache.
\newblock A log-euclidean framework for statistics on diffeomorphisms.
\newblock \emph{Medical Image Computing and Computer-Assisted
  Intervention-MICCAI 2006}, pages 924--931, 2006.

\bibitem[Bilgel et~al.(2016)Bilgel, Prince, Wong, Resnick, and
  Jedynak]{bilgel_multivariate_2016}
M.~Bilgel, J.~L. Prince, D.~F. Wong, S.~M. Resnick, and B.~M. Jedynak.
\newblock A multivariate nonlinear mixed effects model for longitudinal image
  analysis: {Application} to amyloid imaging.
\newblock \emph{NeuroImage}, 134:\penalty0 658--670, July 2016.
\newblock ISSN 10538119.
\newblock \doi{10.1016/j.neuroimage.2016.04.001}.
\newblock URL
  \url{http://linkinghub.elsevier.com/retrieve/pii/S1053811916300349}.

\bibitem[Blezek and Miller(2007)]{blezek_atlas_2007}
D.~J. Blezek and J.~V. Miller.
\newblock Atlas stratification.
\newblock \emph{Medical Image Analysis}, 11\penalty0 (5):\penalty0 443--457,
  Oct. 2007.
\newblock ISSN 13618415.
\newblock \doi{10.1016/j.media.2007.07.001}.
\newblock URL
  \url{https://linkinghub.elsevier.com/retrieve/pii/S1361841507000722}.

\bibitem[Bossa et~al.(2007)Bossa, Hernandez, and
  Olmos]{bossa_contributions_2007}
M.~Bossa, M.~Hernandez, and S.~Olmos.
\newblock Contributions to 3d diffeomorphic atlas estimation: application to
  brain images.
\newblock \emph{Medical Image Computing and Computer-Assisted
  Intervention-MICCAI 2007}, pages 667--674, 2007.

\bibitem[Carmichael et~al.(2013)Carmichael, McLaren, Tommet, Mungas, and
  Jones]{carmichael_coevolution_2013}
O.~Carmichael, D.~G. McLaren, D.~Tommet, D.~Mungas, and R.~N. Jones.
\newblock Coevolution of brain structures in amnestic mild cognitive
  impairment.
\newblock \emph{NeuroImage}, 66:\penalty0 449 -- 456, 2013.
\newblock ISSN 1053-8119.
\newblock \doi{https://doi.org/10.1016/j.neuroimage.2012.10.029}.
\newblock URL
  \url{http://www.sciencedirect.com/science/article/pii/S1053811912010324}.

\bibitem[Cash et~al.(2015)Cash, Frost, Iheme, {\"U}nay, Kandemir, Fripp,
  Salvado, Bourgeat, Reuter, Fischl, Lorenzi, Frisoni, Pennec, Pierson, Gunter,
  Senjem, Jack, Guizard, Fonov, Collins, Modat, Cardoso, Leung, Wang, Das,
  Yushkevich, Malone, Fox, Schott, and Ourselin]{cash_assessing_2015}
D.~M. Cash, C.~Frost, L.~O. Iheme, D.~{\"U}nay, M.~Kandemir, J.~Fripp,
  O.~Salvado, P.~Bourgeat, M.~Reuter, B.~Fischl, M.~Lorenzi, G.~B. Frisoni,
  X.~Pennec, R.~K. Pierson, J.~L. Gunter, M.~L. Senjem, C.~R. Jack, N.~Guizard,
  V.~S. Fonov, D.~L. Collins, M.~Modat, M.~J. Cardoso, K.~K. Leung, H.~Wang,
  S.~R. Das, P.~A. Yushkevich, I.~B. Malone, N.~C. Fox, J.~M. Schott, and
  S.~Ourselin.
\newblock {Assessing atrophy measurement techniques in dementia: Results from
  the MIRIAD atrophy challenge}.
\newblock \emph{{NeuroImage}}, 123:\penalty0 149--164, Dec. 2015.
\newblock \doi{10.1016/j.neuroimage.2015.07.087}.

\bibitem[Christensen et~al.(1994)Christensen, Rabbitt, and
  Miller]{christensen_3d_1994}
G.~E. Christensen, R.~D. Rabbitt, and M.~I. Miller.
\newblock 3d brain mapping using a deformable neuroanatomy.
\newblock \emph{Physics in medicine and biology}, 39\penalty0 (3):\penalty0
  609, 1994.

\bibitem[Cole et~al.(2018)Cole, Ritchie, Bastin, Hern{\'a}ndez, Maniega, Royle,
  Corley, Pattie, Harris, Zhang, et~al.]{cole_brain_2017}
J.~H. Cole, S.~J. Ritchie, M.~E. Bastin, M.~V. Hern{\'a}ndez, S.~M. Maniega,
  N.~Royle, J.~Corley, A.~Pattie, S.~E. Harris, Q.~Zhang, et~al.
\newblock Brain age predicts mortality.
\newblock \emph{Molecular psychiatry}, 23\penalty0 (5):\penalty0 1385, 2018.
\newblock \doi{10.1038/mp.2017.62}.
\newblock URL \url{http://www.nature.com/doifinder/10.1038/mp.2017.62}.

\bibitem[Cury et~al.(2016)Cury, Lorenzi, Cash, Nicholas, Routier, Rohrer,
  Ourselin, Durrleman, and Modat]{cury_spatio-temporal_2016}
C.~Cury, M.~Lorenzi, D.~Cash, J.~M. Nicholas, A.~Routier, J.~Rohrer,
  S.~Ourselin, S.~Durrleman, and M.~Modat.
\newblock Spatio-temporal shape analysis of cross-sectional data for detection
  of early changes in neurodegenerative disease.
\newblock In \emph{International {Workshop} on {Spectral} and {Shape}
  {Analysis} in {Medical} {Imaging}}, pages 63--75. Springer, 2016.

\bibitem[Davatzikos et~al.(2009)Davatzikos, Xu, An, Fan, and
  Resnick]{davatzikos_longitudinal_2009}
C.~Davatzikos, F.~Xu, Y.~An, Y.~Fan, and S.~M. Resnick.
\newblock Longitudinal progression of alzheimer's-like patterns of atrophy in
  normal older adults: the spare-ad index.
\newblock \emph{Brain}, 132\penalty0 (8):\penalty0 2026--2035, 2009.

\bibitem[DeCarli et~al.(2005)DeCarli, Massaro, Harvey, Hald, Tullberg, Au,
  Beiser, D’Agostino, and Wolf]{carli_measures_2005}
C.~DeCarli, J.~Massaro, D.~Harvey, J.~Hald, M.~Tullberg, R.~Au, A.~Beiser,
  R.~D’Agostino, and P.~A. Wolf.
\newblock Measures of brain morphology and infarction in the framingham heart
  study: establishing what is normal.
\newblock \emph{Neurobiology of aging}, 26\penalty0 (4):\penalty0 491--510,
  2005.

\bibitem[Donohue et~al.(2014)Donohue, Jacqmin-Gadda, Le~Goff, Thomas, Raman,
  Gamst, Beckett, Jack, Weiner, Dartigues, and Aisen]{donohue_estimating_2014}
M.~C. Donohue, H.~Jacqmin-Gadda, M.~Le~Goff, R.~G. Thomas, R.~Raman, A.~C.
  Gamst, L.~A. Beckett, C.~R. Jack, M.~W. Weiner, J.-F. Dartigues, and P.~S.
  Aisen.
\newblock Estimating long-term multivariate progression from short-term data.
\newblock \emph{Alzheimer's \& Dementia}, 10\penalty0 (5):\penalty0 S400--S410,
  Oct. 2014.
\newblock ISSN 15525260.
\newblock \doi{10.1016/j.jalz.2013.10.003}.
\newblock URL
  \url{http://linkinghub.elsevier.com/retrieve/pii/S1552526013028732}.

\bibitem[Double et~al.(1996)Double, Halliday, Krill, Harasty, Cullen, Brooks,
  Creasey, and Broe]{double_topography_1996}
K.~Double, G.~Halliday, J.~Krill, J.~Harasty, K.~Cullen, W.~Brooks, H.~Creasey,
  and G.~Broe.
\newblock Topography of brain atrophy during normal aging and alzheimer's
  disease.
\newblock \emph{Neurobiology of Aging}, 17\penalty0 (4):\penalty0 513 -- 521,
  1996.
\newblock ISSN 0197-4580.
\newblock \doi{https://doi.org/10.1016/0197-4580(96)00005-X}.
\newblock URL
  \url{http://www.sciencedirect.com/science/article/pii/019745809600005X}.

\bibitem[Fischl et~al.(2002)Fischl, Salat, Busa, Albert, Dieterich, Haselgrove,
  Van Der~Kouwe, Killiany, Kennedy, Klaveness, et~al.]{fischl2002whole}
B.~Fischl, D.~H. Salat, E.~Busa, M.~Albert, M.~Dieterich, C.~Haselgrove, A.~Van
  Der~Kouwe, R.~Killiany, D.~Kennedy, S.~Klaveness, et~al.
\newblock Whole brain segmentation: automated labeling of neuroanatomical
  structures in the human brain.
\newblock \emph{Neuron}, 33\penalty0 (3):\penalty0 341--355, 2002.

\bibitem[Fjell et~al.(2010)Fjell, Walhovd, Fennema-Notestine, McEvoy, Hagler,
  Holland, Blennow, Brewer, Dale, and the Alzheimer’s Disease
  Neuroimaging~Initiative]{fjell_brain_2010}
A.~M. Fjell, K.~B. Walhovd, C.~Fennema-Notestine, L.~K. McEvoy, D.~J. Hagler,
  D.~Holland, K.~Blennow, J.~B. Brewer, A.~M. Dale, and the Alzheimer’s
  Disease Neuroimaging~Initiative.
\newblock Brain atrophy in healthy aging is related to csf levels of ab1-42.
\newblock \emph{Cerebral Cortex}, 20\penalty0 (9):\penalty0 2069--2079, 2010.
\newblock \doi{10.1093/cercor/bhp279}.
\newblock URL \url{http://dx.doi.org/10.1093/cercor/bhp279}.

\bibitem[Fonteijn et~al.(2012)Fonteijn, Modat, Clarkson, Barnes, Lehmann,
  Hobbs, Scahill, Tabrizi, Ourselin, Fox, and
  Alexander]{fonteijn_event-based_2012}
H.~M. Fonteijn, M.~Modat, M.~J. Clarkson, J.~Barnes, M.~Lehmann, N.~Z. Hobbs,
  R.~I. Scahill, S.~J. Tabrizi, S.~Ourselin, N.~C. Fox, and D.~C. Alexander.
\newblock An event-based model for disease progression and its application in
  familial {Alzheimer}'s disease and {Huntington}'s disease.
\newblock \emph{NeuroImage}, 60\penalty0 (3):\penalty0 1880--1889, Apr. 2012.
\newblock ISSN 10538119.
\newblock \doi{10.1016/j.neuroimage.2012.01.062}.
\newblock URL
  \url{http://linkinghub.elsevier.com/retrieve/pii/S1053811912000791}.

\bibitem[Franke and Gaser(2012)]{franke_longitudinal_2012}
K.~Franke and C.~Gaser.
\newblock Longitudinal {Changes} in {Individual} \textit{{BrainAGE}} in
  {Healthy} {Aging}, {Mild} {Cognitive} {Impairment}, and {Alzheimer}’s
  {Disease}.
\newblock \emph{GeroPsych}, 25\penalty0 (4):\penalty0 235--245, Jan. 2012.
\newblock ISSN 1662-9647, 1662-971X.
\newblock \doi{10.1024/1662-9647/a000074}.
\newblock URL
  \url{http://econtent.hogrefe.com/doi/abs/10.1024/1662-9647/a000074}.

\bibitem[Franke et~al.(2010)Franke, Ziegler, Kl{\"o}ppel, Gaser, Initiative,
  et~al.]{franke_estimating_2010}
K.~Franke, G.~Ziegler, S.~Kl{\"o}ppel, C.~Gaser, A.~D.~N. Initiative, et~al.
\newblock Estimating the age of healthy subjects from t 1-weighted mri scans
  using kernel methods: Exploring the influence of various parameters.
\newblock \emph{Neuroimage}, 50\penalty0 (3):\penalty0 883--892, 2010.

\bibitem[Good et~al.(2001)Good, Johnsrude, Ashburner, Henson, Friston, and
  Frackowiak]{good_voxel-based_2001}
C.~D. Good, I.~S. Johnsrude, J.~Ashburner, R.~N. Henson, K.~J. Friston, and
  R.~S. Frackowiak.
\newblock A {Voxel}-{Based} {Morphometric} {Study} of {Ageing} in 465 {Normal}
  {Adult} {Human} {Brains}.
\newblock \emph{NeuroImage}, 14\penalty0 (1):\penalty0 21--36, July 2001.
\newblock ISSN 10538119.
\newblock \doi{10.1006/nimg.2001.0786}.
\newblock URL
  \url{http://linkinghub.elsevier.com/retrieve/pii/S1053811901907864}.

\bibitem[Guimond et~al.(2000)Guimond, Meunier, and
  Thirion]{guimond_average_2000}
A.~Guimond, J.~Meunier, and J.-P. Thirion.
\newblock Average brain models: A convergence study.
\newblock \emph{Computer vision and image understanding}, 77\penalty0
  (2):\penalty0 192--210, 2000.

\bibitem[Hadj-Hamou(2016)]{hadj-hamou_beyond_2016}
M.~Hadj-Hamou.
\newblock \emph{Beyond volumetry in longitudinal deformation-based morphometry:
  application to sexual dimorphism during adolescence}.
\newblock PhD thesis, Université Côte d'Azur, 2016.
\newblock URL \url{https://tel.archives-ouvertes.fr/tel-01416569/}.

\bibitem[Hadj-Hamou et~al.(2016)Hadj-Hamou, Lorenzi, Ayache, and
  Pennec]{hadj-hamou_longitudinal_2016}
M.~Hadj-Hamou, M.~Lorenzi, N.~Ayache, and X.~Pennec.
\newblock Longitudinal analysis of image time series with diffeomorphic
  deformations: a computational framework based on stationary velocity fields.
\newblock \emph{Frontiers in neuroscience}, 10:\penalty0 236, 2016.
\newblock URL \url{https://www.ncbi.nlm.nih.gov/pmc/articles/PMC4891339/}.

\bibitem[Hutton et~al.(2009)Hutton, Draganski, Ashburner, and
  Weiskopf]{hutton_comparison_2009}
C.~Hutton, B.~Draganski, J.~Ashburner, and N.~Weiskopf.
\newblock A comparison between voxel-based cortical thickness and voxel-based
  morphometry in normal aging.
\newblock \emph{NeuroImage}, 48\penalty0 (2):\penalty0 371--380, Nov. 2009.
\newblock ISSN 10538119.
\newblock \doi{10.1016/j.neuroimage.2009.06.043}.
\newblock URL
  \url{https://linkinghub.elsevier.com/retrieve/pii/S105381190900679X}.

\bibitem[Khanal et~al.(2016)Khanal, Lorenzi, Ayache, and
  Pennec]{khanal_biophysical_2016}
B.~Khanal, M.~Lorenzi, N.~Ayache, and X.~Pennec.
\newblock {A biophysical model of brain deformation to simulate and analyze
  longitudinal MRIs of patients with Alzheimer's disease}.
\newblock \emph{{NeuroImage}}, pages 35--52, July 2016.
\newblock \doi{10.1016/j.neuroimage.2016.03.061}.
\newblock URL \url{https://hal.inria.fr/hal-01305755}.

\bibitem[Kl{\"o}ppel et~al.(2012)Kl{\"o}ppel, Abdulkadir, Jack, Koutsouleris,
  Mour{\~a}o-Miranda, and Vemuri]{kloppel_diagnostic_2012}
S.~Kl{\"o}ppel, A.~Abdulkadir, C.~R. Jack, N.~Koutsouleris,
  J.~Mour{\~a}o-Miranda, and P.~Vemuri.
\newblock Diagnostic neuroimaging across diseases.
\newblock \emph{Neuroimage}, 61\penalty0 (2):\penalty0 457--463, 2012.

\bibitem[Koval et~al.(2017)Koval, Schiratti, Routier, Bacci, Colliot,
  Allassonni{\`e}re, Durrleman, Initiative, et~al.]{koval_statistical_2017}
I.~Koval, J.-B. Schiratti, A.~Routier, M.~Bacci, O.~Colliot,
  S.~Allassonni{\`e}re, S.~Durrleman, A.~D.~N. Initiative, et~al.
\newblock Statistical learning of spatiotemporal patterns from longitudinal
  manifold-valued networks.
\newblock In \emph{International Conference on Medical Image Computing and
  Computer-Assisted Intervention}, pages 451--459. Springer, 2017.

\bibitem[Long et~al.(2012)Long, Liao, Jiang, Liang, Qiu, and
  Zhang]{long_healthy_2012}
X.~Long, W.~Liao, C.~Jiang, D.~Liang, B.~Qiu, and L.~Zhang.
\newblock Healthy aging: an automatic analysis of global and regional
  morphological alterations of human brain.
\newblock \emph{Academic radiology}, 19\penalty0 (7):\penalty0 785--793, 2012.

\bibitem[Lorenzi and Pennec(2013)]{lorenzi_geodesics_2013}
M.~Lorenzi and X.~Pennec.
\newblock Geodesics, parallel transport \& one-parameter subgroups for
  diffeomorphic image registration.
\newblock \emph{International journal of computer vision}, 105\penalty0
  (2):\penalty0 111--127, 2013.
\newblock URL \url{http://link.springer.com/article/10.1007/s11263-012-0598-4}.

\bibitem[Lorenzi et~al.(2013)Lorenzi, Ayache, Frisoni, and
  Pennec]{lorenzi_lcc-demons:_2013}
M.~Lorenzi, N.~Ayache, G.~B. Frisoni, and X.~Pennec.
\newblock {LCC}-{Demons}: a robust and accurate symmetric diffeomorphic
  registration algorithm.
\newblock \emph{NeuroImage}, 81:\penalty0 470--483, 2013.
\newblock URL
  \url{http://www.sciencedirect.com/science/article/pii/S1053811913004825}.

\bibitem[Lorenzi et~al.(2015)Lorenzi, Pennec, Frisoni, Ayache, Initiative, and
  {others}]{lorenzi_disentangling_2015}
M.~Lorenzi, X.~Pennec, G.~B. Frisoni, N.~Ayache, A.~D.~N. Initiative, and
  {others}.
\newblock Disentangling normal aging from {Alzheimer}'s disease in structural
  magnetic resonance images.
\newblock \emph{Neurobiology of aging}, 36:\penalty0 S42--S52, 2015.
\newblock URL
  \url{http://www.sciencedirect.com/science/article/pii/S0197458014005594}.

\bibitem[Lorenzi et~al.(2019)Lorenzi, Filippone, Frisoni, Alexander, and
  Ourselin]{lorenzi2017probabilistic}
M.~Lorenzi, M.~Filippone, G.~B. Frisoni, D.~C. Alexander, and S.~Ourselin.
\newblock Probabilistic disease progression modeling to characterize diagnostic
  uncertainty: Application to staging and prediction in alzheimer's disease.
\newblock \emph{NeuroImage}, 190:\penalty0 56 -- 68, 2019.
\newblock ISSN 1053-8119.
\newblock \doi{https://doi.org/10.1016/j.neuroimage.2017.08.059}.
\newblock URL
  \url{http://www.sciencedirect.com/science/article/pii/S1053811917307061}.
\newblock Mapping diseased brains.

\bibitem[Marchewka et~al.(2014)Marchewka, Kherif, Krueger, Grabowska,
  Frackowiak, Draganski, and Initiative]{marchewka_influence_2013}
A.~Marchewka, F.~Kherif, G.~Krueger, A.~Grabowska, R.~Frackowiak, B.~Draganski,
  and T.~A. D.~N. Initiative.
\newblock Influence of magnetic field strength and image registration strategy
  on voxel-based morphometry in a study of alzheimer's disease.
\newblock \emph{Human Brain Mapping}, 35\penalty0 (5):\penalty0 1865--1874,
  2014.
\newblock \doi{10.1002/hbm.22297}.
\newblock URL \url{https://onlinelibrary.wiley.com/doi/abs/10.1002/hbm.22297}.

\bibitem[Medvedev(1990)]{medvedev_attempt_1990}
Z.~A. Medvedev.
\newblock An attempt at a rational classification of theories of ageing.
\newblock \emph{Biological Reviews}, 65\penalty0 (3):\penalty0 375--398, 1990.

\bibitem[Muralidharan et~al.(2016)Muralidharan, Fishbaugh, Kim, Johnson,
  Paulsen, Gerig, and Fletcher]{muralidharan_bayesian_2016}
P.~Muralidharan, J.~Fishbaugh, E.~Y. Kim, H.~J. Johnson, J.~S. Paulsen,
  G.~Gerig, and P.~T. Fletcher.
\newblock Bayesian covariate selection in mixed-effects models for longitudinal
  shape analysis.
\newblock In \emph{2016 {IEEE} 13th {International} {Symposium} on {Biomedical}
  {Imaging} ({ISBI})}, pages 656--659, Prague, Czech Republic, Apr. 2016. IEEE.
\newblock ISBN 978-1-4799-2349-6.
\newblock \doi{10.1109/ISBI.2016.7493352}.
\newblock URL \url{http://ieeexplore.ieee.org/document/7493352/}.

\bibitem[Ohnishi et~al.(2001)Ohnishi, Matsuda, Tabira, Asada, and
  Uno]{ohnishi2001changes}
T.~Ohnishi, H.~Matsuda, T.~Tabira, T.~Asada, and M.~Uno.
\newblock Changes in brain morphology in alzheimer disease and normal aging: is
  alzheimer disease an exaggerated aging process?
\newblock \emph{American Journal of Neuroradiology}, 22\penalty0 (9):\penalty0
  1680--1685, 2001.

\bibitem[Park and Reuter-Lorenz(2009)]{park_adaptive_2009}
D.~C. Park and P.~Reuter-Lorenz.
\newblock The adaptive brain: aging and neurocognitive scaffolding.
\newblock \emph{Annual review of psychology}, 60:\penalty0 173--196, 2009.

\bibitem[Pedregosa et~al.(2011)Pedregosa, Varoquaux, Gramfort, Michel, Thirion,
  Grisel, Blondel, Prettenhofer, Weiss, Dubourg, Vanderplas, Passos,
  Cournapeau, Brucher, Perrot, and Duchesnay]{scikit-learn}
F.~Pedregosa, G.~Varoquaux, A.~Gramfort, V.~Michel, B.~Thirion, O.~Grisel,
  M.~Blondel, P.~Prettenhofer, R.~Weiss, V.~Dubourg, J.~Vanderplas, A.~Passos,
  D.~Cournapeau, M.~Brucher, M.~Perrot, and E.~Duchesnay.
\newblock Scikit-learn: Machine learning in {P}ython.
\newblock \emph{Journal of Machine Learning Research}, 12:\penalty0 2825--2830,
  2011.

\bibitem[Pini et~al.(2016)Pini, Pievani, Bocchetta, Altomare, Bosco, Cavedo,
  Galluzzi, Marizzoni, and Frisoni]{pini_brain_2016}
L.~Pini, M.~Pievani, M.~Bocchetta, D.~Altomare, P.~Bosco, E.~Cavedo,
  S.~Galluzzi, M.~Marizzoni, and G.~B. Frisoni.
\newblock Brain atrophy in alzheimer’s disease and aging.
\newblock \emph{Ageing Research Reviews}, 30:\penalty0 25 -- 48, 2016.
\newblock ISSN 1568-1637.
\newblock \doi{https://doi.org/10.1016/j.arr.2016.01.002}.
\newblock URL
  \url{http://www.sciencedirect.com/science/article/pii/S1568163716300022}.
\newblock Brain Imaging and Aging.

\bibitem[Reuter et~al.(2012)Reuter, Schmansky, Rosas, and
  Fischl]{reuter_within-subject_2012}
M.~Reuter, N.~J. Schmansky, H.~D. Rosas, and B.~Fischl.
\newblock Within-subject template estimation for unbiased longitudinal image
  analysis.
\newblock \emph{NeuroImage}, 61\penalty0 (4):\penalty0 1402--1418, July 2012.
\newblock ISSN 10538119.
\newblock \doi{10.1016/j.neuroimage.2012.02.084}.
\newblock URL
  \url{https://linkinghub.elsevier.com/retrieve/pii/S1053811912002765}.

\bibitem[Rodrigue and Raz(2004)]{rodrigue2004shrinkage}
K.~M. Rodrigue and N.~Raz.
\newblock Shrinkage of the entorhinal cortex over five years predicts memory
  performance in healthy adults.
\newblock \emph{Journal of Neuroscience}, 24\penalty0 (4):\penalty0 956--963,
  2004.

\bibitem[Rohé et~al.(2016)Rohé, Sermesant, and Pennec]{rohe_barycentric_2016}
M.-M. Rohé, M.~Sermesant, and X.~Pennec.
\newblock Barycentric {Subspace} {Analysis}: a new {Symmetric} {Group}-wise
  {Paradigm} for {Cardiac} {Motion} {Tracking}.
\newblock In \emph{International {Conference} on {Medical} {Image} {Computing}
  and {Computer}-{Assisted} {Intervention}}, pages 300--307. Springer, 2016.

\bibitem[Rosen et~al.(2003)Rosen, Prull, Gabrieli, Stoub, O'hara, Friedman,
  Yesavage, and deToledo Morrell]{rosen2003differential}
A.~C. Rosen, M.~W. Prull, J.~D. Gabrieli, T.~Stoub, R.~O'hara, L.~Friedman,
  J.~A. Yesavage, and L.~deToledo Morrell.
\newblock Differential associations between entorhinal and hippocampal volumes
  and memory performance in older adults.
\newblock \emph{Behavioral neuroscience}, 117\penalty0 (6):\penalty0 1150,
  2003.

\bibitem[Sabuncu et~al.(2009)Sabuncu, Balci, Shenton, and
  Golland]{sabuncu_image-driven_2009}
M.~Sabuncu, S.~Balci, M.~Shenton, and P.~Golland.
\newblock Image-{Driven} {Population} {Analysis} {Through} {Mixture}
  {Modeling}.
\newblock \emph{IEEE Transactions on Medical Imaging}, 28\penalty0
  (9):\penalty0 1473--1487, Sept. 2009.
\newblock ISSN 0278-0062, 1558-254X.
\newblock \doi{10.1109/TMI.2009.2017942}.
\newblock URL \url{http://ieeexplore.ieee.org/document/4804744/}.

\bibitem[Schiratti et~al.(2017)Schiratti, Allassonni{\`e}re, Colliot, and
  Durrleman]{schiratti_bayesian_2016}
J.-B. Schiratti, S.~Allassonni{\`e}re, O.~Colliot, and S.~Durrleman.
\newblock A bayesian mixed-effects model to learn trajectories of changes from
  repeated manifold-valued observations.
\newblock \emph{The Journal of Machine Learning Research}, 18\penalty0
  (1):\penalty0 4840--4872, 2017.
\newblock URL \url{https://hal.archives-ouvertes.fr/hal-01540367/}.

\bibitem[Schmitter et~al.(2015)Schmitter, Roche, Maréchal, Ribes, Abdulkadir,
  Bach-Cuadra, Daducci, Granziera, Klöppel, Maeder, Meuli, and
  Krueger]{schmitter_evaluation_2015}
D.~Schmitter, A.~Roche, B.~Maréchal, D.~Ribes, A.~Abdulkadir, M.~Bach-Cuadra,
  A.~Daducci, C.~Granziera, S.~Klöppel, P.~Maeder, R.~Meuli, and G.~Krueger.
\newblock An evaluation of volume-based morphometry for prediction of mild
  cognitive impairment and {Alzheimer}'s disease.
\newblock \emph{NeuroImage: Clinical}, 7:\penalty0 7--17, 2015.
\newblock ISSN 22131582.
\newblock \doi{10.1016/j.nicl.2014.11.001}.
\newblock URL
  \url{http://linkinghub.elsevier.com/retrieve/pii/S221315821400165X}.

\bibitem[Singh et~al.(2013)Singh, Hinkle, Joshi, and
  Fletcher]{singh_hierarchical_2013}
N.~Singh, J.~Hinkle, S.~Joshi, and P.~T. Fletcher.
\newblock A hierarchical geodesic model for diffeomorphic longitudinal shape
  analysis.
\newblock In \emph{International Conference on Information Processing in
  Medical Imaging}, pages 560--571. Springer, 2013.

\bibitem[Sowell et~al.(2003)Sowell, Peterson, Thompson, Welcome, Henkenius, and
  Toga]{sowell_mapping_2003}
E.~R. Sowell, B.~S. Peterson, P.~M. Thompson, S.~E. Welcome, A.~L. Henkenius,
  and A.~W. Toga.
\newblock Mapping cortical change across the human life span.
\newblock \emph{Nature neuroscience}, 6\penalty0 (3):\penalty0 309--315, 2003.

\bibitem[Tzourio-Mazoyer et~al.(2002)Tzourio-Mazoyer, Landeau, Papathanassiou,
  Crivello, Etard, Delcroix, Mazoyer, and Joliot]{tzourio_automated_2002}
N.~Tzourio-Mazoyer, B.~Landeau, D.~Papathanassiou, F.~Crivello, O.~Etard,
  N.~Delcroix, B.~Mazoyer, and M.~Joliot.
\newblock Automated anatomical labeling of activations in spm using a
  macroscopic anatomical parcellation of the mni mri single-subject brain.
\newblock \emph{Neuroimage}, 15\penalty0 (1):\penalty0 273--289, 2002.

\bibitem[van Velsen et~al.(2013)van Velsen, Vernooij, Vrooman, van~der Lugt,
  Breteler, Hofman, Niessen, and Ikram]{velsen_brain_2013}
E.~F. van Velsen, M.~W. Vernooij, H.~A. Vrooman, A.~van~der Lugt, M.~M.
  Breteler, A.~Hofman, W.~J. Niessen, and M.~A. Ikram.
\newblock Brain cortical thickness in the general elderly population: the
  rotterdam scan study.
\newblock \emph{Neuroscience letters}, 550:\penalty0 189--194, 2013.

\bibitem[Wang et~al.(2007)Wang, Beg, Ratnanather, Ceritoglu, Younes, Morris,
  Csernansky, and Miller]{wang_large_2007}
L.~Wang, F.~Beg, T.~Ratnanather, C.~Ceritoglu, L.~Younes, J.~C. Morris, J.~G.
  Csernansky, and M.~I. Miller.
\newblock Large {Deformation} {Diffeomorphism} and {Momentum} {Based}
  {Hippocampal} {Shape} {Discrimination} in {Dementia} of the {Alzheimer} type.
\newblock \emph{IEEE Transactions on Medical Imaging}, 26\penalty0
  (4):\penalty0 462--470, Apr. 2007.
\newblock ISSN 0278-0062.
\newblock \doi{10.1109/TMI.2006.887380}.
\newblock URL \url{http://ieeexplore.ieee.org/document/4141196/}.

\end{thebibliography}
\addcontentsline{toc}{chapter}{Bibliographie}

\section{Appendix}
\label{sec:appendix}

\subsection{Validation on a synthetic dataset}
\label{subsec:appendix:synthetic}

The regional values of atrophy set are given in Table~\ref{table:atrophy_values}. For each subject, the values are sampled around these means with a 5\% standard deviation.

\begin{table}[ht!]
  \centering
\begin{tabular}{|lll|}
  \hline
  brain area        & mean pathological (in \%) & mean healthy (in \%) \\
  \hline
  white matter      &  1.0  & 0.8  \\
  cortex            &  3.0  & 0.4  \\
  hippocampi        &  5.2  & 1.0  \\
  amygdalae         &  5.2  & 1.0  \\
  entorhinal cortex &  6.5  & 0.7  \\
  temporal poles    &  6.2  & 0.6  \\
  other areas       &  0.0  & 0.0  \\
  \hline
\end{tabular}
\caption{Specified mean regional atrophy for the healthy and the pathological evolutions. The goal is to get simple but realistic atrophy patterns. It should be noted that the atrophy is specified using the divergence of the SVF in the area. The local volume changes are computed using a spatio-temporal integration scheme.}
\label{table:atrophy_values}
\end{table}

\begin{figure}[ht!]
  \centering
  \includegraphics[width=0.48\textwidth]{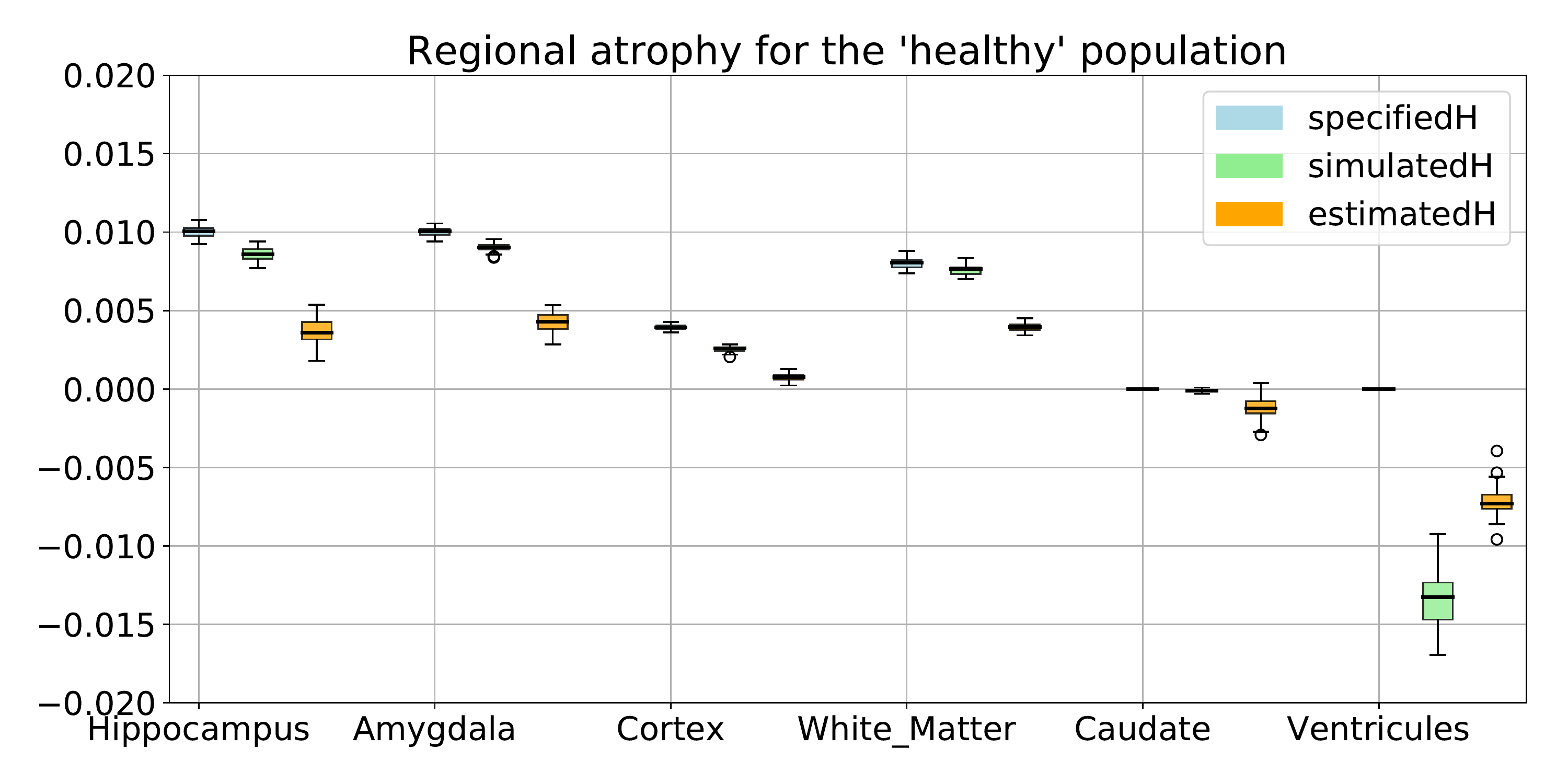}
  \includegraphics[width=0.48\textwidth]{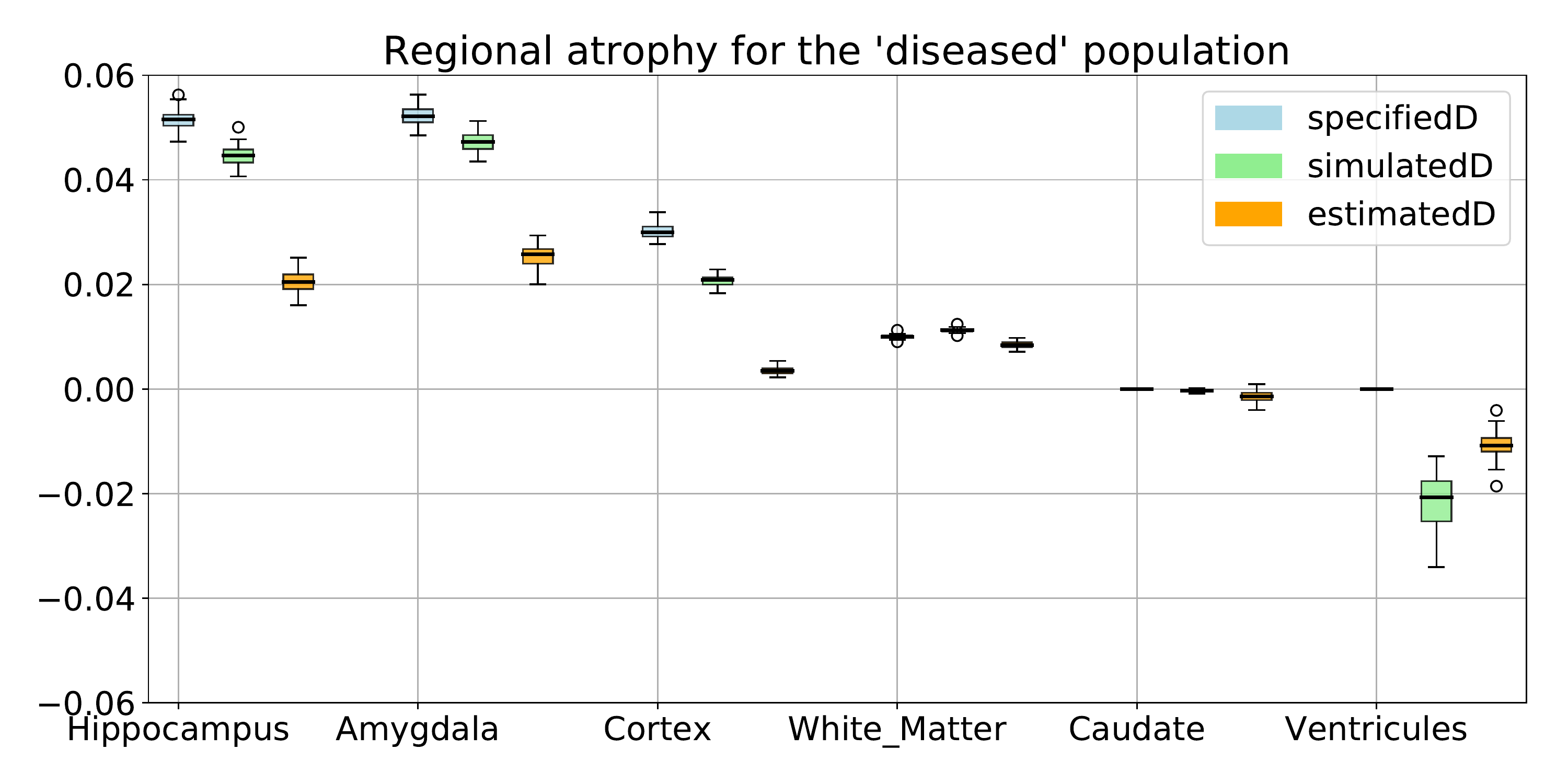}
  \includegraphics[width=0.48\textwidth]{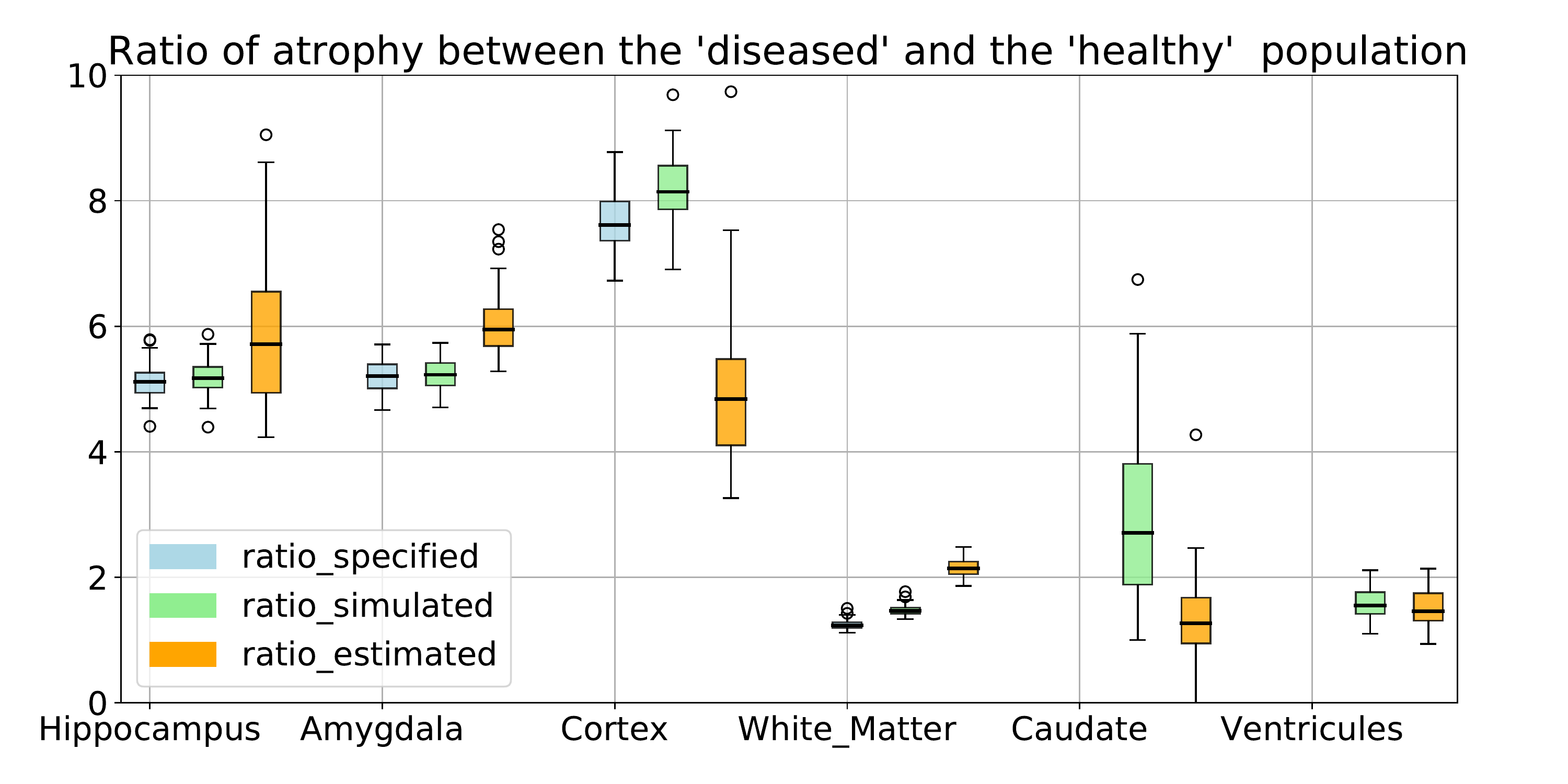}
  \caption{Comparison of prescribed, simulated and estimated atrophy values for several brain regions. The difference between prescription and simulation can be explained by numerical approximations in the biophysical model while the estimation is biased due to spatial regularization. However the estimation bias is consistent accross regions and subjects.}
  \label{fig:synth_atrophy_boxplot}
\end{figure}

The comparison with the values obtained after simulation or estimated through registration are shown in Figure~\ref{fig:synth_atrophy_boxplot}. The estimation can be biased by the spatial regularization and the loss of information in intensity homogeneous areas. The relative changes is however similar between the two populations and the method is by consequence adapted to compare the two evolutions.

\subsection{Bias on the estimated template trajectories}
\label{subsec:appendix:bias}

The norms of the SVFs parametrizing the template trajectories have an effect on the normalization of the individual biomarkers estimation. We estimate here the bias on the the norm of the estimation relatively to the estimated norm.

For the normal aging trajectory we have:
\begin{align*}
  E( \| \hat{\mathrm{v}}_{A} \|^2 ) &=  \frac{1}{|\mathcal{G}_h|^2} E(< \sum_{k \in \mathcal{G}_h}{v^k} | \sum_{k \in \mathcal{G}_h}{v^k} >) \\
  &=  \frac{1}{|\mathcal{G}_h|^2} \sum_{i, j \in \mathcal{G}_h} E(< s_{MA}^i \mathrm{v}_{A} + s_{DS}^i \mathrm{v}_{D} + v_{r}^i | s_{MA}^j \mathrm{v}_{A} + s_{DS}^j \mathrm{v}_{D} + v_{r}^j >) \\
  &=  \frac{1}{|\mathcal{G}_h|^2} \sum_{i, j \in \mathcal{G}_h} E(s_{MA}^i s_{MA}^j) \|\mathrm{v}_{A}\|^2 + E(s_{DS}^i s_{DS}^j) \|\mathrm{v}_{D}\|^2 + E(s_{MA}^i s_{DS}^j + s_{DS}^i s_{MA}^j) < \mathrm{v}_{A} | \mathrm{v}_{D} > + E(< v_{r}^i | v_{r}^j >) \\
\end{align*}

Assuming that the subjects are independent and identically distributed:
 \begin{align*} 
  E( \| \hat{\mathrm{v}}_{A} \|^2 )  &=   \|\mathrm{v}_{A}\|^2 + \frac{1}{|\mathcal{G}_h|^2} \sum_{i \in \mathcal{G}_h} \Var(s_{MA}^i) \|\mathrm{v}_{A}\|^2 + \Var(s_{DS}^i) \|\mathrm{v}_{D}\|^2 + 2 E(s_{MA}^i s_{DS}^i) < \mathrm{v}_{A} | \mathrm{v}_{D} > + E( \| v_{r}^i \|^2) \\
  &=  \left(1 + \frac{\Var(s_{MA}^h)}{|\mathcal{G}_h|} \right) \|\mathrm{v}_{A}\|^2 + \frac{\Var(s_{DS}^h)}{|\mathcal{G}_h|} \|\mathrm{v}_{D}\|^2 + \frac{E( \| v_{r}^h \|^2)}{|\mathcal{G}_h|}  \\
 \end{align*}

And similarly for $\| \hat{\mathrm{v}}_{D} \| $:
\begin{align*}
  E( \| \hat{\mathrm{v}}_{D} \|^2 ) &= E \left( \| \frac{1}{|\mathcal{G}_{ad}|} \sum_{k \in \mathcal{G}_{ad}}{v^k} - \hat{\mathrm{v}}_{A} \|^2 \right) \\
  &= E(\| \hat{\mathrm{v}}_{A} \|^2) + E \left( \| \frac{1}{|\mathcal{G}_{ad}|} \sum_{k \in \mathcal{G}_{ad}}{v^k} \|^2 \right) - 2 E \left( < \frac{1}{|\mathcal{G}_{ad}|} \sum_{k \in \mathcal{G}_{ad}}{v^k} | \hat{\mathrm{v}}_{A} > \right)\\
  &= \left(1 + \frac{\Var(s_{MA}^{ad})}{|\mathcal{G}_{ad}|} \right)  \|\mathrm{v}_{A}\|^2 + \left(1 + \frac{\Var(s_{DS}^{ad})}{|\mathcal{G}_{ad}|} \right) \|\mathrm{v}_{D}\|^2 + 2<\mathrm{v}_{A} | \mathrm{v}_{D}  > + \frac{E( \| v_{r}^{ad} \|^2)}{|\mathcal{G}_{ad}|} \\
  &\phantom{= 1}  + E( \| \hat{\mathrm{v}}_{A} \|^2 ) - 2 (\| \mathrm{v}_{A} \|^2 + <\mathrm{v}_{A} | \mathrm{v}_{D} >) \\
  &= \left(\frac{\Var(s_{MA}^{ad})}{|\mathcal{G}_{ad}|} + \frac{\Var(s_{MA}^{h})}{|\mathcal{G}_h|} \right) \|\mathrm{v}_{A}\|^2 + \left(1 + \frac{\Var(s_{DS}^{ad})}{|\mathcal{G}_{ad}|} + \frac{\Var(s_{DS}^{h})}{|\mathcal{G}_h|} \right) \|\mathrm{v}_{D}\|^2 + \frac{E( \| v_{r}^{ad} \|^2)}{|\mathcal{G}_{ad}|} + \frac{E( \| v_{r}^h \|^2)}{|\mathcal{G}_h|}
\end{align*}

The terms $\Var(s_{MA})$, $\Var(s_{DS})$ and  $\| v_{r} \|^2$ are related to the individual variability that is not modeled and to the noise in the estimated deformations.
We empirically estimate the variances in the training population and we get that, for our training dataset,  $\| \hat{\mathrm{v}}_{A} \|^2 \approx 1.65 \| \mathrm{v}_{A} \|^2$ and $\| \hat{\mathrm{v}}_{D} \|^2 \approx 1.21 \| \mathrm{v}_{D} \|^2$.
We should however note that the same subjects from $\mathcal{G}_h$ are used to estimate the template anatomy $T_0$ and the normal aging template trajectory $\mathrm{v}_{A}$. Therefore, we are certainly underestimating the bias coming from the intra-subject morphological variability for this population.

This bias has a direct influence on the markers estimations. Indeed for a subject $k$, the markers are solution of the linear system~\ref{eqt:subject_parameters} involving $\|\mathrm{v}_{A}\|^2$, $\|\mathrm{v}_{D}\|^2$ for which the biased estimators are used.

\end{document}